\documentclass[a4paper,11pt]{article}
\usepackage{jcappub}
\usepackage[T1]{fontenc}
\usepackage{caption}
\usepackage{subcaption}
\usepackage{comment}
\usepackage{amsmath, amssymb, graphics}
\usepackage{mathtools}
\usepackage{tablefootnote}
\usepackage{soul}
\usepackage[normalem]{ulem}
\usepackage{color}
\usepackage{float}
\usepackage{makecell}
\usepackage{booktabs}
\usepackage[usenames,dvipsnames,svgnames,table]{xcolor}
\usepackage{indentfirst}
\usepackage{leftidx}
\usepackage{multirow}
\definecolor{LightCyan}{rgb}{0.88,1,1}
\newcolumntype{a}{>{\columncolor{LightCyan}}c}

\newcommand{\apjl}{Astrophys. J. Lett.}
\newcommand{\apj}{Astrophys. J.}

\newcommand{\aap}{Astron. Astrophys.}

\newcommand{\mnras}{Mon. Not. Roy. Astron. Soc.}

\newcommand{\jcap}{JCAP}
\newcommand{\physrep}{Phys. Reports}

\usepackage{xspace}

\newcommand{\hmpc}{$h^{-1}\mathrm{Mpc}$\xspace}
\newcommand{\hmsun}{$h^{-1}M_\odot$\xspace}

\title{Cross-correlations of the Cosmic Neutrino Background with the Dark Matter field: HR-DEMNUni simulation analysis}

\author[1]{Beatriz Hern\'andez-Molinero,}
\author[2,3]{Matteo Calabrese,}
\author[2]{Carmelita Carbone,}
\author[4]{Alessandro Greco,}
\author[5,6]{Raul Jimenez,}
\author[1,7]{Carlos Pe\~na Garay,}

\affiliation[1]{Laboratorio Subterr\'aneo de Canfranc, 22880 - Canfranc-Estaci\'on, Huesca, Spain.}
\affiliation[2]{INAF – Istituto di Astrofisica Spaziale e Fisica cosmica di Milano (IASF-MI), Via Alfonso Corti 12, I-20133 Milano, Italy}
\affiliation[3]{Astronomical Observatory of the Autonomous Region of the Aosta Valley (OAVdA), Loc. Lignan 39, I-11020, Nus (Aosta Valley), Italy}
\affiliation[4]{Department of Astronomy, University of Florida, 211 Bryant Space Science Center, Gainesville, FL 32611, USA}
\affiliation[5]{ICCUB, University of Barcelona, Marti  i Franques 1, E-08028 Barcelona, Spain.}
\affiliation[6]{Instituci\`o Catalana de Recerca i Estudis Avan\c{c}ats, Pg. Lluis Companys 23, Barcelona, E-08010, Spain.}
\affiliation[7]{I2SysBio, CSIC-University of Valencia, 46071 - Valencia, Spain.}

\emailAdd{bhernandez@lsc-canfranc.es}
\emailAdd{alessandro.greco.1@phd.unipd.it}
\emailAdd{calabrese@oavda.it}
\emailAdd{carmelita.carbone@inaf.it}
\emailAdd{raul.jimenez@icc.ub.edu}
\emailAdd{cpenya@lsc-canfranc.es}

\abstract{We use the high-resolution HR-DEMNUni simulations to compute cross-correlations of the Cosmic Neutrino Background quantities, like neutrino density, deflection angle, and velocity, with  other quantities, like cold dark matter density and effective weak lensing convergence, by accounting for the space-time delay between signals on Earth. We provide this to theoretically illustrate how much can be learned from these cross-correlation signals, once the cosmic neutrino background is detected with instruments in multiple locations. Against a naive expectation of null cross-correlation, we show that the signal is non zero, specially at the largest scales. We also discuss the scenario of co-located cross-correlations between dark matter, weak lensing and a future neutrino-induced photon emission signal. As cross-correlations will be comparable to auto-correlations of the cosmic neutrino background itself and are less affected by cosmic variance and shotnoise, these might be the ones to be measured first. Our predictions thus provide the imprint of what cosmological massive neutrinos, with total mass $\sum{m_\nu} \sim 0.1$ eV, should look like from cosmological observations.}
\begin{document}
\maketitle	

\section{Introduction}
\label{sec:Introduction}
The Cosmic Neutrino Background (C${\nu}$B), which carries information about the Universe from just one second after the Big Bang, has not yet been directly detected. However, its eventual detection would provide invaluable insights into both the early stages of the Universe and the nature of neutrinos \cite{Dolgov:2002wy,Dolgov_2008,Long_2014}. Some experiments are being designed and developed to achieve this goal, including tabletop experiments based on tritium capture~\cite{PTolemy}. In addition to direct detection, C${\nu}$B neutrinos may be probed through their subtle effects on dark matter lensing~\cite{Pacos,Hernandez-Molinero_2024}, producing an observable feature in the matter profile of massive galaxy clusters. Furthermore, their interaction with astrophysical objects containing nuclear-density matter, such as neutron stars and planets, could lead to the emission of photons at keV frequencies~\cite{NS}, which may be detectable from Earth. 

The study of neutrino cosmological signals is motivated by the current limits provided by cosmological surveys, which indicate that the sum of neutrino masses is approaching the minimum total mass inferred by oscillation data (about $0.059$ eV) \cite{DESI}. This strongly suggests, within the framework of Bayesian evidence, that the hierarchy is normal \cite{Fer1, Fer2}. If this is the case, neutrinoless double beta decay experiments \cite{Goeppert-Mayer, furry}, which are the current experimental method to discover the nature of neutrinos, will need to operate in the several-ton range, which presents significant challenges. Given this scenario, cosmological signals emerge as a promising way to reveal the nature of neutrinos.

In previous work~\cite{Hernandez-Molinero:2022zoo,Hernandez-Molinero_2024} we showed that gravitational effects can modify the helicity distribution of cosmic background neutrinos, thus altering the expected capture rates in neutrino detectors.
Such experiments could reveal the fundamental nature of neutrinos, specifically, whether they are Dirac or Majorana particles, by measuring the average capture rate. If neutrinos are Dirac particles, only left-handed states will be detected; however, if they are Majorana, both left- and right-handed neutrinos contribute to the signal~\cite{Long_2014,Roulet:2018fyh,Hernandez-Molinero:2022zoo}.
Given the importance of such a possible discovery, it is worth exploring in detail the possible angular dependence of this signal.

In this paper, in order to quantify the angular dependence of cosmic neutrino background signals,  we have produced full-sky maps of Cold Dark Matter (CDM) density, neutrino density, neutrino velocity, Weak Lensing (WL)  accounting for the contribution from CDM and neutrinos, and neutrino deflection angle along with all their cross-correlations. We will demonstrate the angular dependence of these signals and identify where they are maximal. This, in turn, will provide clues for discovering the nature of neutrinos from observations of the sky.

Correlations of large-scale structure and neutrino signals have been studied before. Ref.~\cite{Michney} used analytical considerations to show that the C$_{\nu}$B exhibits a Sachs-Wolfe effects. Ref.~\cite{Hannestad} solved the Boltzman equation to compute the anisotropy of the C$_{\nu}$B in linear theory, including the effect of gravitational lensing in neutrinos of masses $0.1$ eV. Ref.~\cite{Tully} computes the anisotropies in the C$_{\nu}$B using Boltzmann codes to solve its hierarchy and exploring several line-of-sight integrations, finding measurable anisotropies. Finally, Ref.~\cite{Elbers} used a constrained numerical simulation of dark matter within $200$ h$^{-1}$ Mpc of the Milky Way to trace the trajectories of neutrinos as they passed through the dark matter towards the Earth.

The paper is organized as follows. In \autoref{sec:Methodology} we present the HR-DEMNUni simulations used to produce the full sky maps, for both CDM and neutrinos, which are shown in \autoref{sec:Maps}. All cross-correlations and autocorrelation power spectra are gathered in \autoref{sec:Cross-correlations} where we explore two cases, first the direct neutrino detection in \autoref{sec:direct detection}, and second, the case of co-located cross-correlations in \autoref{sec:Co-location}. Finally, the discussion and conclusions are addressed in \autoref{sec:Conclusions}.

\section{Methodology}
\label{sec:Methodology}

In previous work \cite{Hernandez-Molinero:2022zoo,Hernandez-Molinero_2024}, we demonstrated that gravity modifies the helicity content of the cosmic neutrino background: neutrinos are deflected by collapsed structures, resulting in helicity flips. These changes can be quantified through the deflection angle accumulated along neutrino trajectories since the formation of large-scale structures. Building on our earlier studies~\cite{Hernandez-Molinero:2022zoo,Hernandez-Molinero_2024}, the present work extends the analysis to the full sky, enabling us to explore the cross-correlations between this signal and the clustering distributions of CDM and cosmic neutrinos.

All maps used in this study have been generated using the ``Dark Energy and Massive Neutrino Universe'' (DEMNUni) suite of large N-body simulations~\cite{carbone_2016}. The DEMNUni simulations were produced to investigate the Large-Scale Structure (LSS) of the Universe in the presence of massive neutrinos and dynamical Dark Energy (DE), and were conceived for the nonlinear analysis and modeling of different probes, including CDM, halo, and galaxy clustering~\cite{castorina_2015,moresco_2016,zennaro_2018,ruggeri_2018,bel_2019,parimbelli_2021,parimbelli_2022,Guidi_2022, Baratta_2022, Gouyou_Beauchamps_2023, Hernandez_2024b, Verdiani_2025, SHAM-Carella_in_prep}, weak lensing, CMB lensing, Sunyaev-Zeldovich and integrated Sachs-Wolfe effects ~\cite{carbone_2016,roncarelli_2015,fabbian_2018, Ingoglia_2024, Hernandez-Molinero_2024, Luchina_etal_2025}, cosmic void statistics~\cite{kreisch_2019,schuster_2019,Verza_2019,Verza_2022a, Verza_2022b,Verza_2024}, as well as cross-correlations among these probes~\cite{Vielzeuf_2022,Cuozzo2022}.

In particular, we have used the new high-resolution (HR) simulations with 64 times better mass resolution than previous standard DEMNUni runs: the HR-DEMNUni simulations are characterized by a comoving volume of $(500 \: h^{-1}\mathrm{Mpc})^3$ filled with $2048^3$ CDM particles and, when present, $2048^3$ neutrino particles with total neutrino mass $M_\nu \equiv \sum m_\nu =0.16\, {\rm eV}$ considered in the degenerate mass scenario (the three active neutrinos in HR-DEMNUni being represented by one single particle with mass $M_\nu$). The simulations are initialised at $z_{\rm in}=99$ with Zeldovich initial conditions. The initial power spectrum is rescaled to the initial redshift using the rescaling method developed in~\cite{zennaro_2017}. The initial conditions are then generated with a modified version of the \texttt{N-GenIC} software, assuming Rayleigh random amplitudes and uniform random phases. The other cosmological parameters of the simulations are based on a Planck 2013~\cite{planck2013} $\Lambda$CDM reference cosmology (with massless neutrinos), in particular: $n_{\rm s}=0.96$, $A_{\rm s}=2.1265 \times 10^{-9}$, $h=H_0/[100\, 
{\rm km} \, s^{-1}{\rm Mpc}^{-1}]=0.67$, $\Omega_{\rm b}=0.05$, and $\Omega_{\rm m}=\Omega_{\rm CDM} + \Omega_{\rm b} + \Omega_\nu =0.32$; $H_0$ is the Hubble constant at the present time, $n_{\rm s}$ is the spectral index of the initial scalar perturbations, $A_{\rm s}$ is the scalar amplitude, $\Omega_{\rm b}$ the baryon density parameter, $\Omega_{\rm m}$ is the total matter density parameter, $\Omega_{\rm CDM}$ the CDM density parameter, and $\Omega_\nu$ the neutrino density parameter. In the presence of massive neutrinos, $\Omega_{\rm b}$ and $\Omega_{\rm m}$ are kept fixed to the above values, while $\Omega_{\rm CDM}$ is changed accordingly. \autoref{tab:neutrino_params} summarises the masses of CDM and neutrino particles together with the neutrino fraction $f_\nu \equiv \Omega_\nu / \Omega_{\rm m}$.

As mentioned earlier, using these simulations, we have computed several full-sky maps of key quantities: CDM density contrast, neutrino density contrast, neutrino velocity field, cosine of the neutrino deflection angle ($\cos{\theta_{\rm \nu}}$) and WL convergence. In particular, to obtain the full-shy map of $\cos{\theta_{\rm \nu}}$, the deflection angle for each neutrino in the simulation is calculated as the change in its velocity vector from $z=3$ down to $z=0$, following the same method as in Ref.~\cite{Hernandez-Molinero:2022zoo,Hernandez-Molinero_2024}.  

\begin{table}[t]
\centering
\vspace{2ex}
\setlength{\tabcolsep}{0.7em}
\begin{tabular}{cccc}
\toprule
\multicolumn{4}{c}{HR-DEMNUni} \\
\midrule
$\sum m_\nu$  [eV] & $f_\nu$ & $m_{\rm p}^{\rm CDM}$  [\hmsun] & $m_{\rm p}^\nu$  [\hmsun] \\
\hline
0  & 0 & $1.2921\times 10^{9}$ & $0$ \\
\hline
0.16  & 0.012 & $1.2767\times 10^{9}$ & $1.5441\times 10^7$ \\
\bottomrule
\end{tabular}
\caption{
Summary of particle masses and neutrino fractions implemented in the HR-DEMNUni simulations. The first column shows the total neutrino mass, the second the fraction of neutrinos and matter density parameters, and the last two columns show the corresponding mass of CDM and neutrino particles implemented in the simulations. 
}
\label{tab:neutrino_params}
\end{table}

\section{Full-sky Cold Dark Matter and Neutrino maps}
\label{sec:Maps}
The CDM and neutrino maps are obtained from the corresponding particle distributions of the HR-DEMNUni simulations at the corresponding redshift\footnote{As discussed in detail later in the manuscript, this study makes use of the snapshot outputs at redshifts $z=0$ and $z=0.06$ from the HR-DEMNUni simulation.} where the cross-correlation is performed. 
With the observer placed at the center, we project all particles inside a radius of $250-260 \text{ Mpc} h^{-1}$, depending on the redshift of the considered snapshot to construct a surface mass density $\Sigma$ for each 3D sphere of $\nu$ and CDM. All particles distributed within each of these 3D matter shells are projected onto 2D spherical maps, assigning a specific sky pixel to each particle via the \texttt{HEALpix}\footnote{\url{http://healpix.sourceforge.net}}~\cite{Gorski} pixelization procedure. For each pixel of the \texttt{HEALpix} grid - or position $\boldsymbol{\theta}$ in the sky - one has the following expression:  
\begin{equation}
\Sigma(\boldsymbol{\theta}) = \dfrac{n \cdot m_p}{A_{\rm pix}}\,,
\label{eq:surfmass}
\end{equation}
where $n$ is the number of particles in the pixel, $A_{\rm pix}$ is the pixel area in steradians, and $m_p$ is the particle mass of the $N$-body simulation in each pixel, $m_{\rm p}^{\rm CDM}$ or $m_{\rm p}^\nu$. 

We independently produced full-sky surface mass density maps for CDM and neutrino particles, on a \texttt{HEALpix} grid with $n_\text{side} = 4096$, which corresponds to a pixel resolution of $0.85$ arcmin. For neutrinos and CDM densities, we use the snapshot at $z=0.06$ in the case of neutrino direct detection, and the snapshot at $z=0$ in the case of indirect detection, as described in \autoref{sec:Cross-correlations}. In the latter case we use the position of the neutrino and CDM particles to construct corresponding density maps at $z=0$ alone, which are the ones considered for the cross-correlation with WL maps in the case of indirect neutrino detection~\footnote{For the maximum distance considered ($250-260$ Mpc $h^{-1}$) we treat the neutrino particle as it was never deflected again, see~\cite{Hernandez-Molinero_2024}. We showed that the last deflection for neutrinos with low masses takes place at about 200 Mpc $h^{-1}$, where the most massive Lainakea-like (supercluster) structure occurs. 
Therefore, limiting our analysis to $z = 0$, we are confident that this assumption is valid and no errors are introduced. Instead, we use the snapshot at $z=0.06$ for the case of neutrino direct detection. From this snapshot we construct two different shells of CDM and neutrinos, separated by a distance such that they are simultaneously} received from the observer at $z=0$. 
If $n$ particles fall into a particular pixel, the total mass in it is $n \cdot m_p$.
A similar method has been applied to build the catalogue for the neutrino velocity modulus and deflection angle, where the velocity and deflection angle have been calculated for each neutrino particle of the simulation; given the particle position in the snapshot of the simulation, full-sky maps of neutrino velocity and deflection angle have been produced, representing the distribution of those quantities on a 2D \texttt{HEALpix} grid. In this case, the value returned in the full-sky map is the average over all particle velocities, or deflection angles, that fall into that pixel.
\begin{figure}[t!]
    \centering
    \begin{tabular}{cc}
    \includegraphics[width=0.49\textwidth]{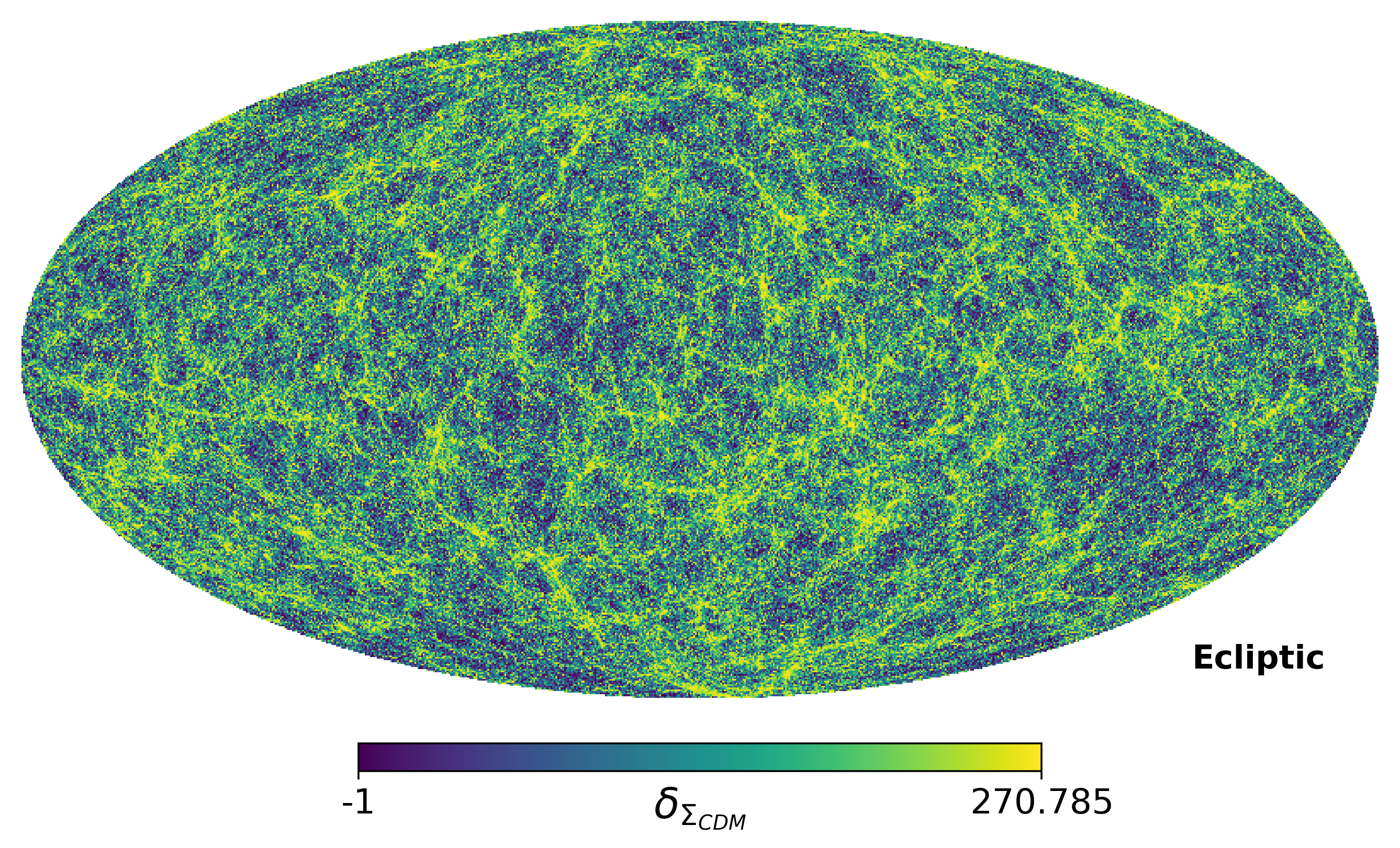} & \includegraphics[width=0.49\textwidth]{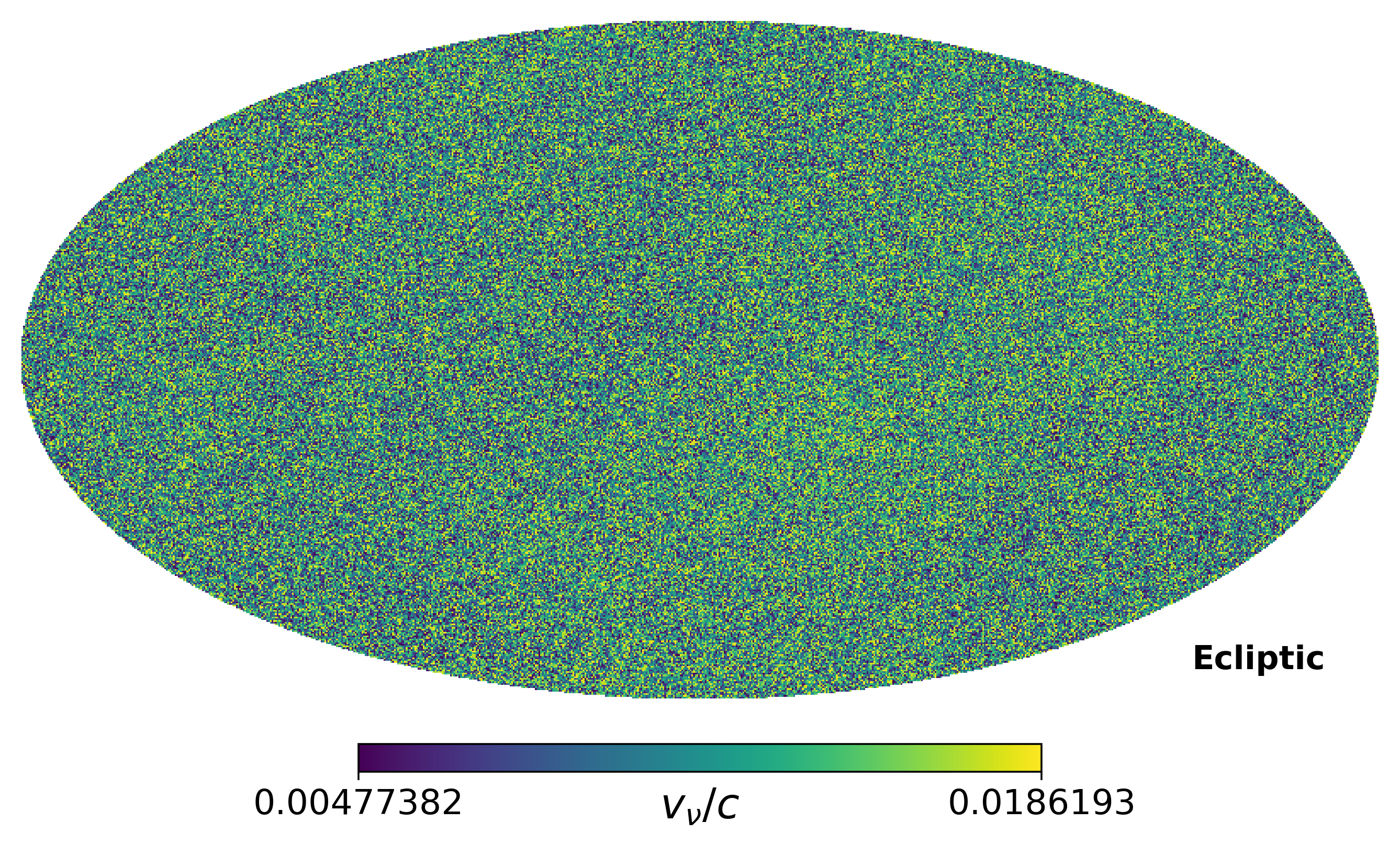} \\
    \includegraphics[width=0.49\textwidth]{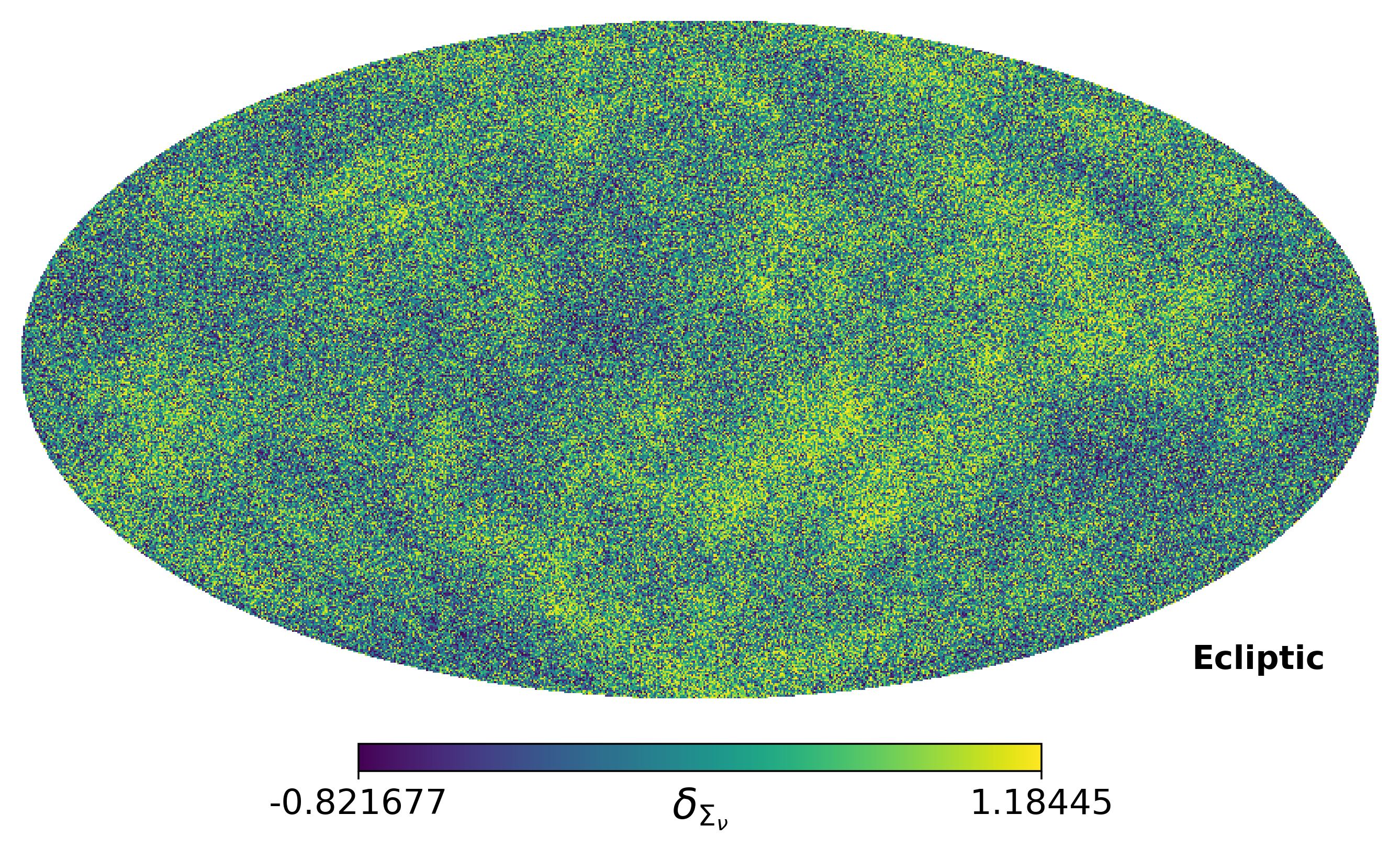} & \includegraphics[width=0.49\textwidth]{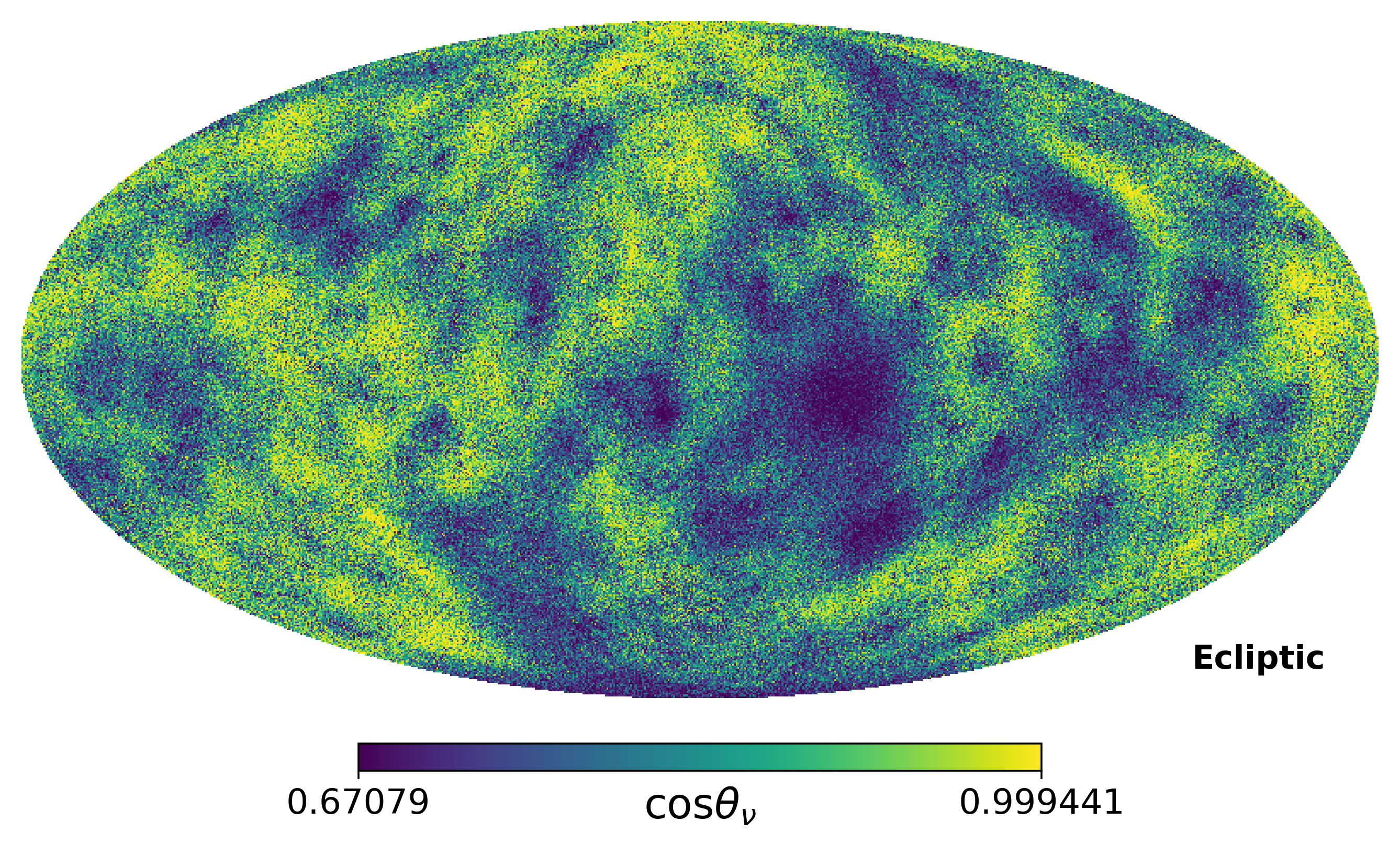}
    \end{tabular}
    \caption{Full-sky map of 250\hmpc comoving radius around an observer placed at the centre of CDM surface overdensity (top left),  neutrino velocity modulus (top right), neutrino surface overdensity (bottom left), and neutrino mean cosine of deflection angle (bottom right).}
    \label{fig:maps_0}
\end{figure}

The corresponding maps for each of the above quantities are shown in the four panels of \autoref{fig:maps_0} when we use the $z=0$ snapshot alone.
It should be noted that in the left panels of \autoref{fig:maps_0} the displayed quantities are not the surface densities but $\delta_\Sigma$, where $\delta_\Sigma\equiv(\Sigma-\bar{\Sigma})/\bar{\Sigma}$ is the density contrast, i.e. the difference in surface density of each pixel with respect to the mean surface density of all pixels, divided by the latter.

Observing the left panels of \autoref{fig:maps_0}, i.e. the distribution of $\delta_{\rm \Sigma_{CDM}}$ and $\delta_{\rm \Sigma_\nu}$, one can notice that the clustering of neutrinos follows the clustering of CDM, as neutrinos fall into the CDM potential wells. But, due to their hot thermal velocities, especially for low neutrino masses such as $\Sigma m_\nu = 0.16$ eV, neutrino perturbations have lower amplitude than CDM ones, and, compared to CDM perturbations, in the neutrino case the small scale clustering is erased, since neutrinos cluster only above their free streaming length, $\lambda_{\rm FS}$. Therefore, the $\delta_{\rm \Sigma_\nu}$ distribution appears much more diffused and smoother (as also the minimum and maximum values in the color bars show) than the $\delta_{\rm \Sigma_{CDM}}$ distribution, but in any case their correlation is observable from the maps. Instead, comparing the left and right bottom panels of \autoref{fig:maps_0}, i.e. the distributions of $\delta_{\rm \Sigma_\nu}$ and of $\cos{\theta}_{\rm \nu}$, it looks like the two quantities are anticorrelated on large scales, to more dense ``brighter'' regions in the neutrino overdensity, there correspond ``darker'' regions where $\cos{\theta}_{\rm \nu}$ assumes smaller values. Since regions with a lower mean $\cos{\theta}_{\rm \nu}$ correspond to larger angles of deflection, and since, as noted above, $\delta_{\rm \Sigma_\nu}$ is correlated to $\delta_{\rm \Sigma_{CDM}}$, this means that, in regions where the CDM overdensity is higher, the gravitational force will cause neutrinos to deflect more.
 Finally, in the top right panel of \autoref{fig:maps_0}, it is difficult to identify any structure in the distribution of the neutrino velocity modulus, because the contribution of coherent neutrino velocities is very small and hidden by their thermal velocity, which is dominant especially in the low neutrino masses considered.

\section{Power Spectra and Cross-Correlations}
\label{sec:Cross-correlations}

Two approaches will be discussed below. First, we consider the case of direct detection of cosmic neutrinos with good energy resolution at various locations on Earth and their cross-correlation with the photon signals of the CDM structures. In this scenario, we should cross-correlate photons with neutrinos located at different distances from the observer position to reach it at the same time, given their different speeds. Second, we will consider the co-located scenario, where neutrinos at their positions produce or induce the emission of photons that are detected on Earth at the same time as photons from weak-lensing (WL) or CMB observables. In this case, we suppose that the distribution of free-streaming neutrinos produces anisotropies in the pattern of neutrino-induced photons (in the same way as CDM perturbations induce anisotropies in CMB photons) and that somehow we can reconstruct the evolving neutrino field from them.

\subsection{Direct Detection: Cross-correlations of  CDM and Neutrinos detected on Earth}
\label{sec:direct detection}

Let us consider a future scenario in which neutrinos are directly detected on Earth using a set of detectors, potentially Ptolemy-like, though not necessarily limited to that design, while CDM continues to be traced, as usual, by the light emitted from galaxies residing within dark matter halos. In this case, any cross-correlation with LSS must account for a space-time delay between the arrival of neutrinos and photons. Since photons travel faster than neutrinos, a neutrino detected on Earth at a given time could not have originated from the same location as a simultaneously arriving photon. To enable meaningful cross-correlations, one must instead consider photons carrying information from more distant locations,  such that both reach Earth at the same time.

To model this space-time offset, we compare CDM maps and neutrino maps generated using the procedure described in \autoref{sec:Maps}. As mentioned above and explained in \autoref{sec:past_neutrino_methods}, for simplicity we consider a single snapshot approach, at $z=0.06$ in the case of direct detection. This redshift corresponds to a maximum comoving distance from the observer of $r \simeq 260 \,h^{-1}$ Mpc. For CDM, we then select particles located within a shell centred on this distance, in a radius $240 \lesssim r \lesssim 260\,h^{-1}$Mpc, and for neutrinos we consider a shell of radius $10 \lesssim r \lesssim 50\,h^{-1}$Mpc. Although a more complete treatment would involve summing over multiple neutrino-CDM shell pairs with appropriate time delays, the specific configuration used here serves as a representative example.

The shell selection is based on the neutrino mass of the simulation: neutrinos have individual masses of $m_\nu=(0.16\,{\rm eV})/3=0.053$ eV, corresponding to typical thermal velocities of $0.1-0.2\,c$. This implies that photons from a CDM shell at distance $\sim 250\,h^{-1}$ Mpc will reach the observer at the same time as neutrinos located at $25-50\,h^{-1}$ Mpc from Earth. Although the full signal would arise from a continuous distribution of neutrinos across all distances, weighted by the Fermi-Dirac distribution, our choice effectively captures the dominant contribution to the cross-correlation.

\begin{figure}
    \centering
    \includegraphics[width=\linewidth]{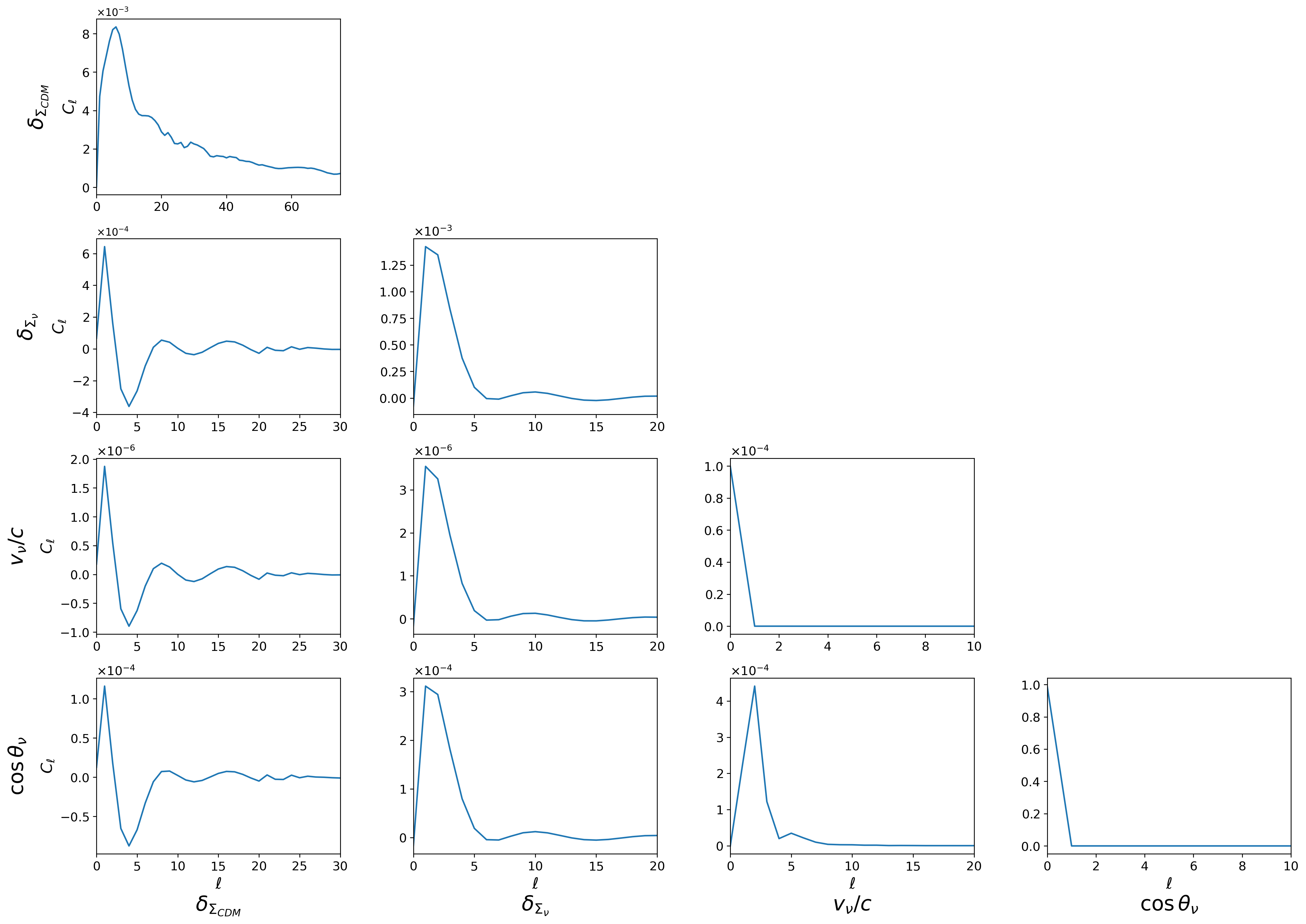}
    \caption{Cross-correlations power spectra (off-diagonal panels) along with autocorrelation power spectra (diagonal panels) of CDM and neutrino maps. The CDM density contrast ($\delta_{\rm \Sigma_{\rm CDM}}$) map is constructed from a shell at a radius of $250\,h^{-1}$Mpc from an observer on the Earth, while the neutrino density contrast ($\delta_{\rm \Sigma_\nu}$), neutrino velocity ($v_{\rm\nu}/c$) and cosine of the neutrino deflection angle ($\cos\theta_{\rm\nu}$) are obtained from a shell at a radius of $30\,h^{-1}$Mpc from the same observer.}
    \label{fig:cls_onearth}
\end{figure}

The cross-correlation power spectra between $\delta_{\rm \Sigma_{\rm CDM}}$, $\delta_{\rm \Sigma_\nu}$, $v_{\rm\nu}/c$ and $\cos\theta_{\rm\nu}$ for the direct neutrino detection case are shown in \autoref{fig:cls_onearth} (off-diagonal panels).  These angular correlations in the spherical harmonic domain have been obtained by applying the \texttt{anafast} routine from the \texttt{HEALpix} package~\cite{Gorski} to the corresponding maps.  Since we work with a single cosmological realization and aim to reduce statistical noise, the resulting power spectra were smoothed using the \texttt{Savgol} routine from the \texttt{scipy} Python package\footnote{{\url{https://docs.scipy.org/doc/scipy/reference/generated/scipy.signal.savgol\_filter.html}}}.

Let us begin by analysing the diagonal panels that show the autocorrelations. Focusing first on the neutrino properties, we observe that their autocorrelation spectra peak at smaller multipoles $\ell$ compared to that of CDM, as expected. This behaviour is driven by the low mass of neutrinos ($\Sigma m_\nu = 0.16$ eV) which limits their ability to cluster on small scales. Since neutrinos decoupled just one second after the Big Bang, they have streamed freely ever since, retaining an almost frozen Fermi-Dirac momentum distribution. Their velocities are tied to this distribution via their mass, with lower masses corresponding to higher thermal dispersion. As a result, neutrinos cluster only on large scales, which is reflected in the observed peak: $\ell\sim3$ for $\delta_{\rm \Sigma_\nu}$ and $\ell<2$ for both $v_{\rm \nu}/c$ and $\cos{\theta}_{\rm \nu}$. This is consistent with what has been found in~\cite{Hernandez-Molinero_2024}, where it has been shown that the deflection reaches its maximum at the super-cluster scales. In contrast, the CDM density contrast auto-correlation spectrum, $\delta_{\rm \Sigma_{CDM}}$, peaks at higher multipoles, $\ell\sim10$, reflecting its ability to cluster on smaller scales, again as expected. 

Turning now to the neutrino-neutrino cross-correlations, we observe that $\delta_{\rm \Sigma_\nu}$ peaks at the same angular scale when cross-correlated with both $v_{\rm \nu}/c$ and $\cos{\theta}_{\rm \nu}$, specifically at $\ell\sim3$. In particular, although they peak on the same scale, the cross-correlation between $\delta_{\rm \Sigma_{\nu}}$ and $\cos{\theta_{\rm \nu}}$ is stronger than that between $\delta_{\rm \Sigma_{\nu}}$ and $v_{\nu}/c$. Furthermore, the cross-correlation between $v_{\nu}/c$ and $\cos{\theta_{\rm \nu}}$ also peaks at $\ell\sim3$, and its amplitude is comparable to that of the cross-correlation between $\delta_{\rm \Sigma_{\nu}}$ and $\cos{\theta_{\rm \nu}}$.

The remaining cross-correlations involve the CDM density contrast, which deserves particular attention. One may naively expect that the delayed pairing between neutrino- and CDM-shells will result in no cross-correlation. However, we show that this is not the case, particularly on the largest scales. In fact, the neutrino- and CDM-shells are separated by a distance of about $200\,h^{-1}$Mpc, which in 3D-space would correspond to $k \sim 0.03\,h{\rm Mpc}^{-1}$, a scale where the 3D cross-power spectrum between neutrinos and CDM can be computed and is non-vanishing. Moreover, even if here we use the same comoving snapshot at $z=0.06$ both for CDM and neutrinos (for which a shell centred at $r=30\,h^{-1}$Mpc along the line-of-sight should actually correspond to $z \sim 0.006$, i.e. one order of magnitude lower than $z=0.06$), the large-scale structure does not evolve much in this short time, as shown in \autoref{fig:comparison}, although it is still fully nonlinear at most scales. \autoref{fig:cls_onearth} shows that the resulting cross-correlations are non-negligible. In fact, the signal of the cross-correlation between $\Sigma_{\rm CDM}$ and the  neutrino velocity $v_\nu$  is of the same order of magnitude as in the co-located case shown in \autoref{fig:cls_colocated}, although the signal is significant only on the largest scales. In agreement with previous results, the cross-correlations of the neutrino deflection angle are greater than those of the neutrino velocity field.\footnote{To obtain directional information, several detectors need to be installed on Earth as it is done for gravitational waves.}

Moreover, since CDM and neutrino maps are constructed from different radial shells (the CDM shell lies at $r \simeq 250\,h^{-1}$ Mpc, whereas the neutrino shell is centered at $r \simeq 30\,h^{-1}$Mpc), this means that the same multipole $\ell$ corresponds to different physical scales in each case. With this in mind, we examine the first column of \autoref{fig:cls_onearth}, which shows the cross-correlations between CDM and each neutrino observable. A consistent pattern emerges: a maximum correlation at $\ell\sim2$ and a maximum anticorrelation (that is, minimum correlation) at $\ell\sim4$. Furthermore, the cross-correlations with $\delta_{\rm \Sigma_{\nu}}$ and $\cos{\theta_{\rm \nu}}$ are similar in amplitude, whereas the cross-correlation with $v_{\rm \nu}/c$ is about two orders of magnitude smaller, as expected. Given the disparity in physical scales, the cross-correlation  is dominated by the contribution of positive overdensities for dark matter and neutrinos at $\ell\sim2$, while the cross-correlation is dominated by the contribution of positive overdensities for neutrinos with negative overdensities from the tail of the dark matter profiles at $\ell\sim4$.

\subsection{Indirect Detection: Co-located Cross-correlations via Neutrino-induced Photon Emission}
\label{sec:Co-location}

\begin{figure}
    \centering
    \includegraphics[width=0.8\textwidth]{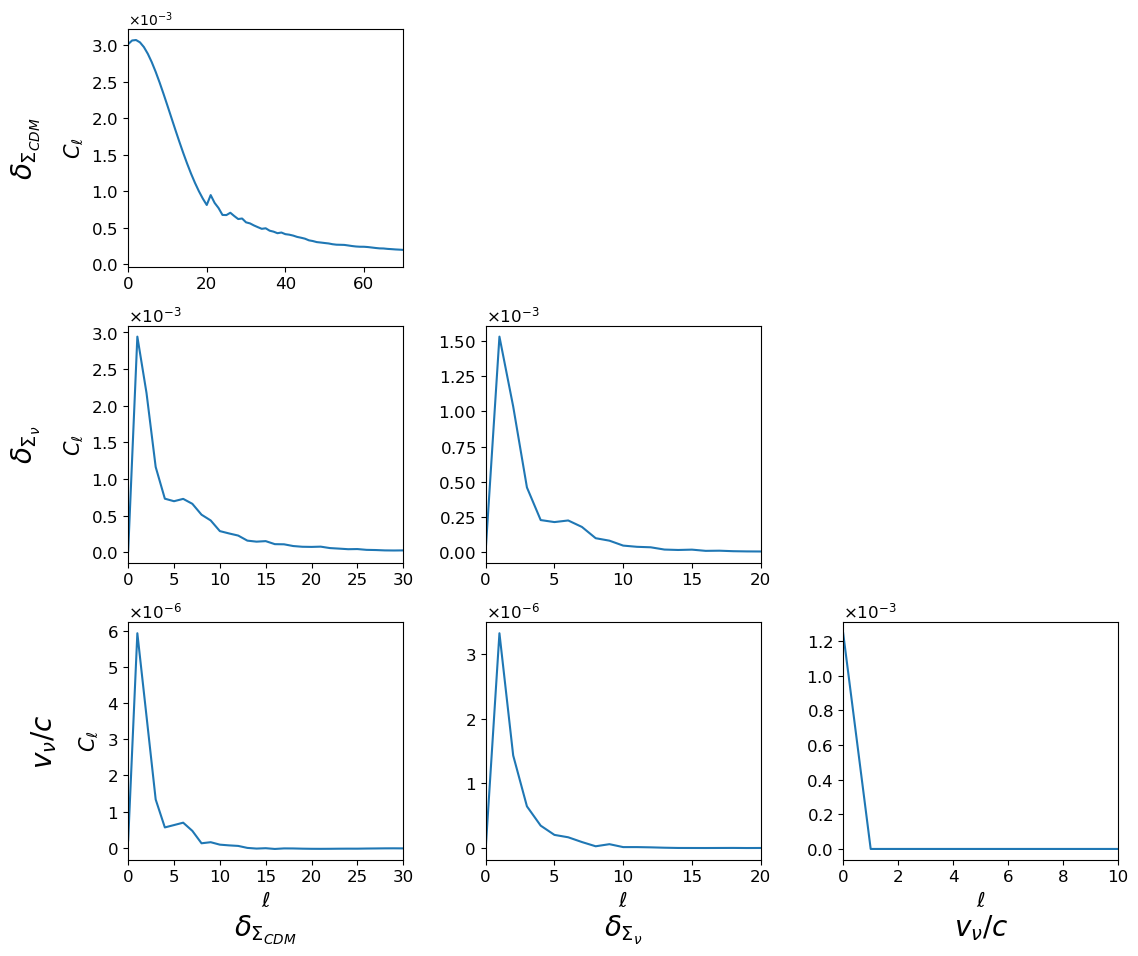}
    \caption{Cross-correlation power spectra (off-diagonal panels) along with autocorrelation power spectra (diagonal panels) for the maps shown in \autoref{fig:maps_0}, excluding the $\cos\theta_\nu$ map due to the loss of directionality in neutrino-induced photon emission processes. Unlike the previous case, these spectra are computed using maps constructed from the entire simulation box at $z=0$, rather than from radial shells.}
    \label{fig:cls_colocated}
\end{figure}

The observables discussed in the previous section require the challenging detection of very low-energy cosmological neutrinos in multiple locations. In this section, we explore an alternative route in the hypothetical case mentioned in \autoref{sec:Cross-correlations}, i.e. that the light signal induced by cosmological neutrinos capture could be observed in future telescopes. In this scenario, neutrinos at their positions would produce or induce the emission of photons that are detected on Earth at the same time as photons from galaxies in collapsed CDM structures. Two mechanisms could enable such an emission: (i) neutrino clustering could subtly affect the weak lensing signal, introducing a measurable distortion that is separable from the CDM-only one, thereby allowing a partial reconstruction of the neutrino distribution~\cite{Pacos}, and (ii) neutrino interactions with astrophysical objects that contain nuclear density matter, such as neutron stars or planets~\cite{NS}, might result in photon emission if a neutrino is captured under certain conditions. These photons could then be detected on Earth, enabling cross-correlation measurements with LSS.

In both cases, the photon emission mechanisms erase directional information about individual neutrinos. As a result, for this scenario, we focus only on auto- and cross-correlations involving the CDM and neutrino density contrasts as well as the neutrino velocity field. These are shown in \autoref{fig:cls_colocated}.

The previously discussed thermal distribution of neutrinos, which limits their clustering on small scales, is again visible in these results. The autocorrelation of $\delta_{\rm \Sigma_\nu}$, and its cross-correlation with $\delta_{\rm \Sigma_{CDM}}$, both peak at large scales, that is, $\ell\sim3$. This behaviour is also present in the autocorrelation of the neutrino velocity field $v_{\rm \nu}/c$, which peaks at the same multipole. Similarly to the direct detection case, the cross correlations between $v_{\rm \nu}/c$ and both $\delta_{\rm \Sigma_{CDM}}$ and $\delta_{\rm \Sigma_\nu}$ are of comparable magnitude. As expected, the autocorrelation of the CDM density contrast peaks at smaller scales, corresponding to larger multipoles, with a maximum around $\ell\sim10$. Aside from the cross-correlations involving $\delta_{\rm \Sigma_{CDM}}$, the overall behaviour of the power spectra remains consistent with that of the direct detection scenario discussed earlier. However, when CDM-related cross-correlation power spectra were compared in both scenarios, notable differences emerged. These differences arise because, in the direct case, CDM and neutrinos are sampled from distinct radial shells, whereas in this scenario they are co-located. Consequently, the oscillatory feature observed in the CDM-$\nu$ cross-correlations of \autoref{fig:cls_onearth} is absent in \autoref{fig:cls_colocated}. 

This distinction in clustering behavior between CDM particles and neutrinos is also evident when comparing the sky maps. The $\delta_{\rm \Sigma_{CDM}}$ map (top left panel in \autoref{fig:maps_0}) reveals the typical large-scale structure of the Universe full of vast galaxy filaments and voids spanning incredibly large distances, a direct consequence of the stronger clustering of CDM. In contrast, the $\delta_{\rm \Sigma_\nu}$ map (bottom left panel in \autoref{fig:maps_0}) exhibits a smoother  distribution, consistent with the higher thermal velocities of neutrinos which, via free streaming, suppress their small scale clustering. This visual difference aligns with their respective power spectra, with the CDM peaking at $\ell\sim10$ and the neutrinos peaking at $\ell\sim3$.

\begin{figure}[]
    \centering
    \begin{subfigure}[b]{0.48\linewidth}
        \includegraphics[width=\linewidth]{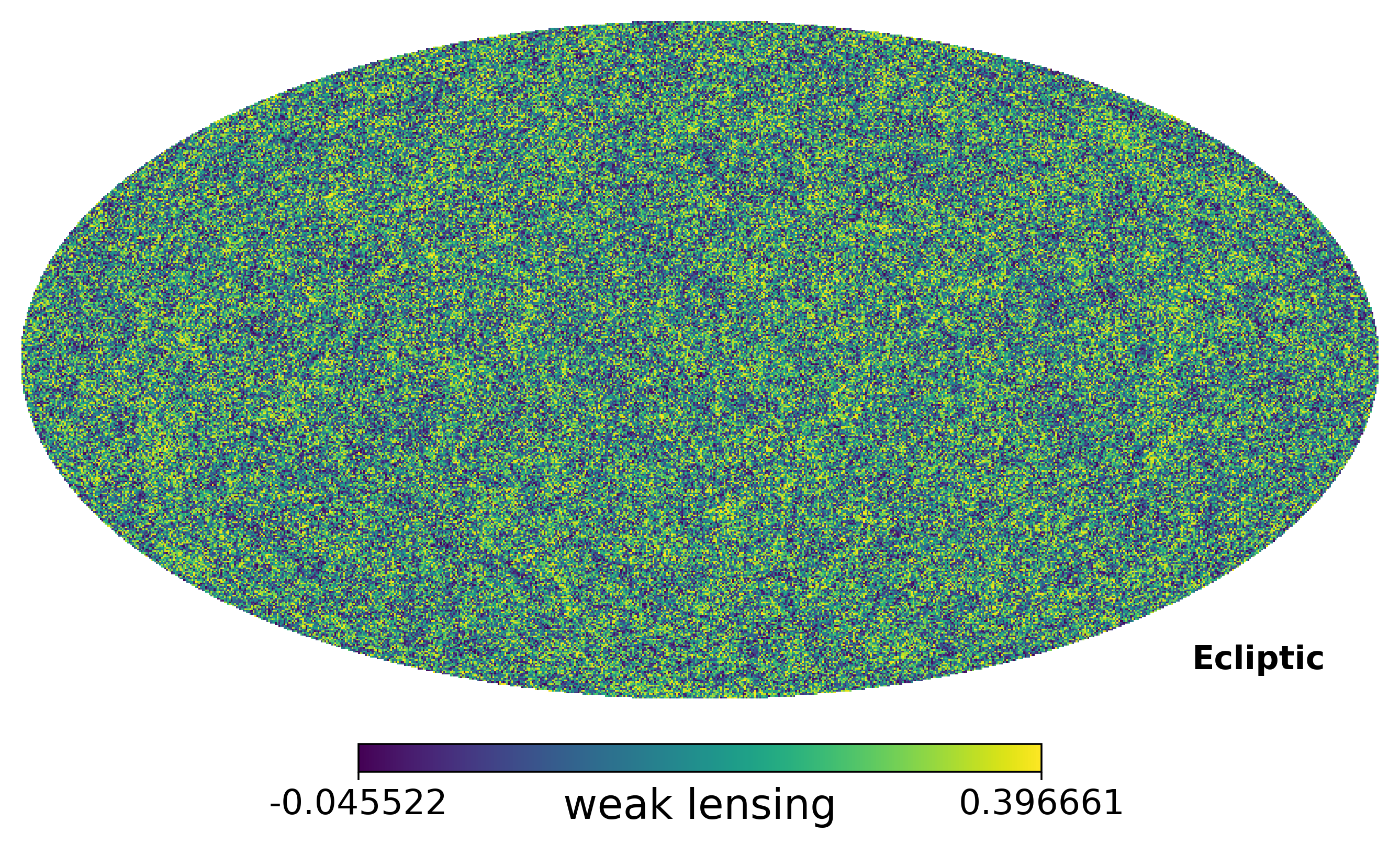}
        \caption{Map}
        \label{fig:CMB_map}
    \end{subfigure}
    \hfill
    \begin{subfigure}[b]{0.48\linewidth}
        \includegraphics[width=\linewidth]{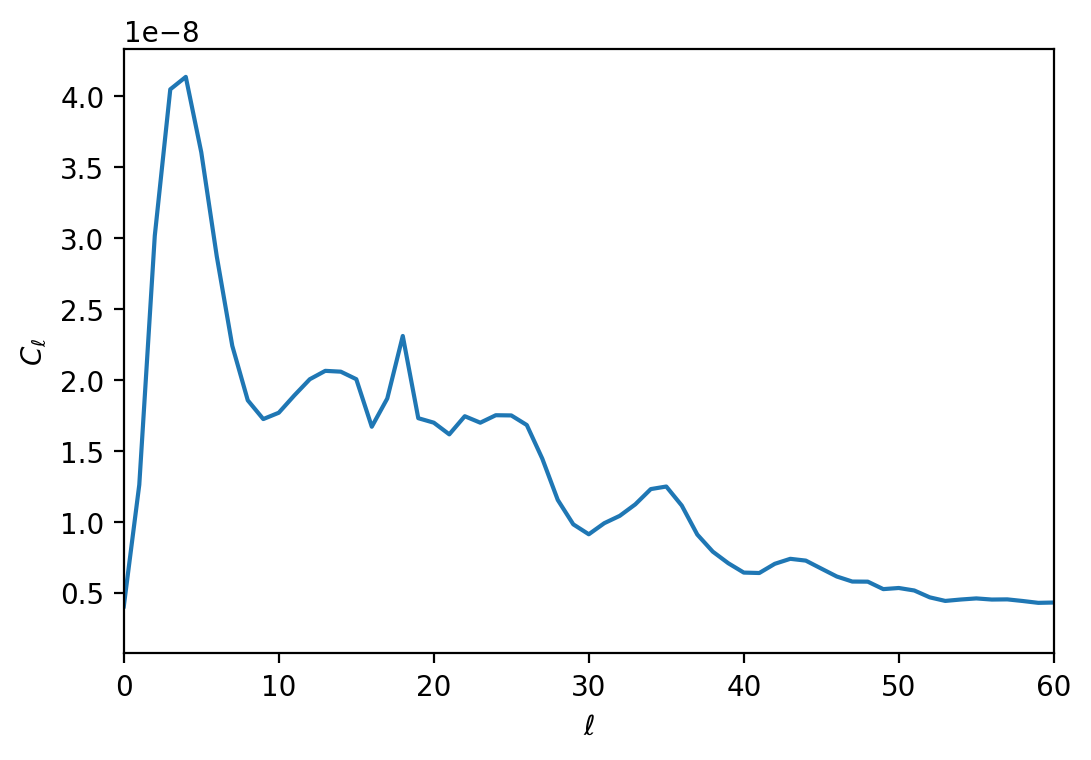}
        \caption{Power spectrum}
        \label{fig:CMB_power_spectrum}
    \end{subfigure}    
    \caption{WL map constructed as described in the text by considering the total contribution of neutrinos and CDM, together with the corresponding total power spectrum.}
\end{figure}

\begin{figure}[b!]
    \centering
    \begin{subfigure}[b]{0.48\linewidth}
        \centering
        \includegraphics[width=\linewidth]{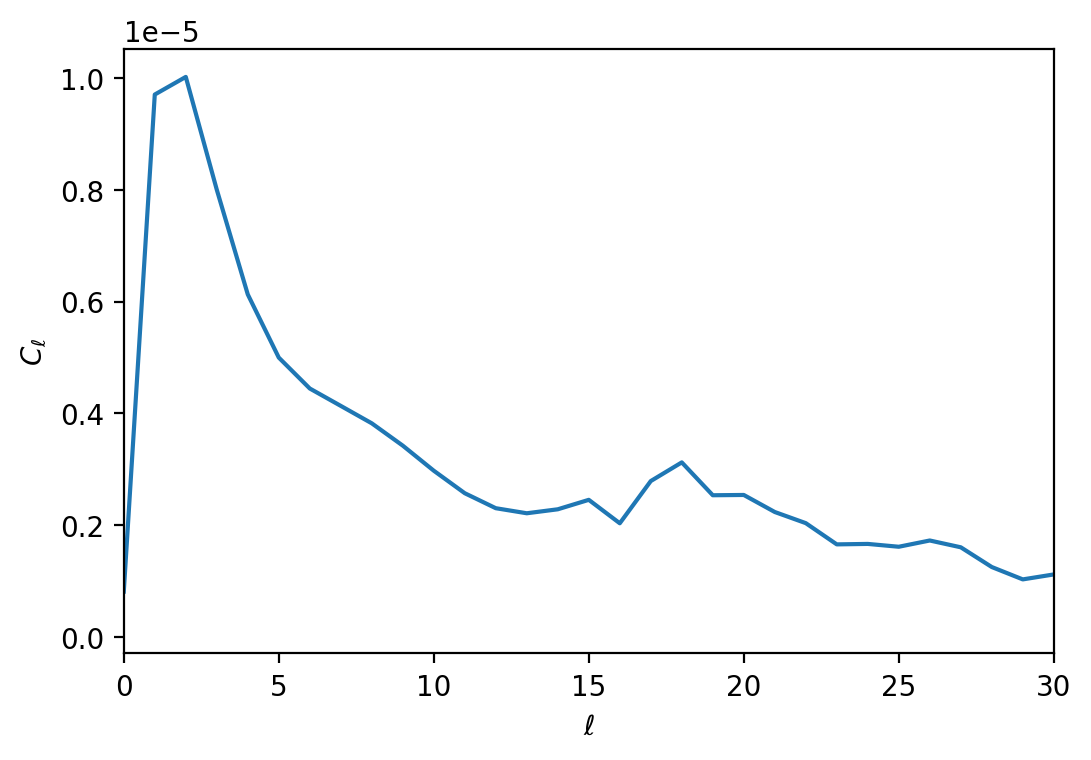}
        \caption{WL cross $\delta_{\rm \Sigma_{\rm CDM}}$}
        \label{fig:CMB_vs_dm_den_CrossPowerSpectrum}
    \end{subfigure}
    \hfill
    \begin{subfigure}[b]{0.48\linewidth}
        \centering
        \includegraphics[width=\linewidth]{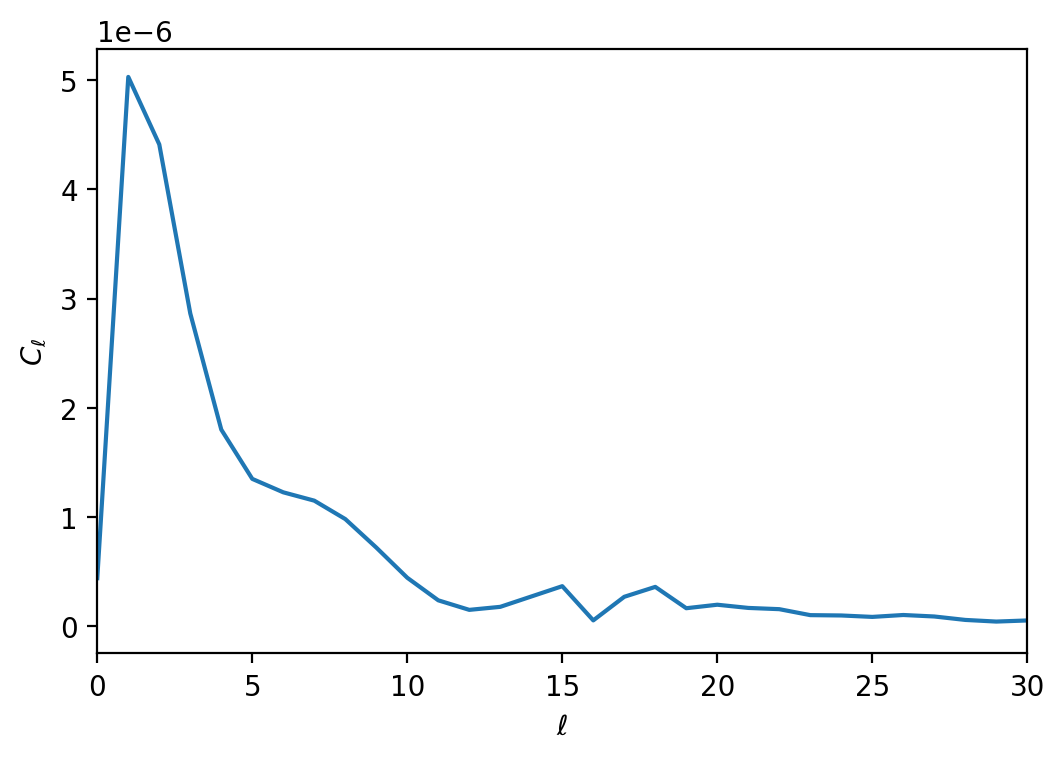}
        \caption{WL cross $\delta_{\rm \Sigma_{\rm \nu}}$.}
        \label{fig:CMB_vs_nu_den_CrossPowerSpectrum}
    \end{subfigure}
        \caption{WL cross-correlation power spectra with CDM and neutrino density contrast. The cross-correlation spectra found in this case shows features similar to the one that cross-correlates the dark matter. This is expected, as the weak lensing convergence is basically a re-weighting of the dark matter distribution given a $n(z)$.}
    
    \label{fig:weaklensingcross}
\end{figure}

Additionally, it is worth exploring in detail the cross-correlation of the neutrino field with LSS by using an actual observable of the latter. In this case, we have chosen to construct a WL convergence map, where the sources are distributed through a distribution of $n(z)$, which peaks at about $z=0.5$. This procedure follows the approaches of~\cite{carbone2009, calabrese2015}, and was developed to perform high-resolution CMB and weak lensing simulations~\cite{fabbian2018, Accuracy_weak_lensing_simulations-Hilbert+20}. The shape of the $n(z)$ and the parameters are taken from the galaxy selection function expected for the photometric \textit{Euclid} survey~\cite{Laureijs_2011, EuclidVII_2020}: 
\begin{equation}
\label{eq:euclid_n}
    n(z) \propto \left(\frac{z}{z_{0}}\right)^{\beta} \exp\Big[ -\Big(\frac{z}{z_{0}} \Big)^{\gamma}\Big]\,,
\end{equation}
where $\beta$ = 2, $\gamma$ = 1.5, $z_{0}$ = $z_{m}/\sqrt{2}$ and $z_{m} = 0.5$. This effective WL map related to a source galaxy distribution can be computed as in~\citep{2001PhR...340..291B} from a full, past-light-cone reconstruction following the procedure described in~\citep{calabrese2015,parimbelli_2022,Cuozzo2022}. Since our maps (neutrino velocities and angles, CDM distributions, etc.) are constructed in a comoving volume of $(500h^{-1}\rm{Mpc})^3$, the WL convergence is an observable that can maximize the cross-correlation with the neutrino fields at $z=0$ rather than, for instance, the CMB lensing potential field, which has a broader kernel in redshift (for the case of photon-emission from neutrinos, one could potentially consider the cross-correlation between WL maps and $\Sigma_\nu$ at larger $z$ as the signal will be larger thanks to the WL geometrical efficiency). The results of the autocorrelation for the WL map are shown in \autoref{fig:CMB_power_spectrum}. As expected, it shows the standard scale dependence for the WL signal~\citep{fabbian_2018,parimbelli_2022,Cuozzo2022}. The cross-correlations with the neutrino density contrast at $z=0$ are shown in \autoref{fig:weaklensingcross} together with the cross-correlation with $\delta_{\rm \Sigma_{CDM}}$. These cross-correlation power spectra peak at similar multipoles as those shown in Figs.~\ref{fig:cls_onearth} and \ref{fig:cls_colocated}, although their amplitudes are smaller than in the previous cases. Notably, the cross-correlations between the WL convergence and both CDM and neutrino density contrasts are stronger than the WL autocorrelation itself. This highlights the cross-correlations between neutrinos and CDM as particularly promising observables for future experimental surveys. 

\section{Conclusions}
\label{sec:Conclusions}

We have cross-correlated the C$\nu$B density contrast, velocity modulus and cosine of deflection angle maps between them and with CDM density contrast and WL convergence maps, in a full-sky picture where the observer is placed at the center. To do so, we have used the new high-resolution DEMNUni simulations that contain $2048^3$ particles of each type (CDM and neutrinos) in a comoving volume of $(500h^{-1}\rm{Mpc})^3$. The simulations used consider a degenerate neutrino mass scenario with three active neutrinos of $\Sigma m_\nu=0.16$~eV. The resulting maps have a pixel resolution of $0.85$ arcmin. To compute the deflection angle of neutrinos, we have used the same recipe as in~\cite{Hernandez-Molinero:2022zoo,Hernandez-Molinero_2024}, that is, calculating the deflection angle as the angle between the velocity vectors of neutrinos at $z=3$ and $z=0$.

Our results are perfectly consistent with the theoretical expectations from our previous analysis~\cite{Hernandez-Molinero:2022zoo,Hernandez-Molinero_2024}. One of these features is that the CDM density contrast power spectrum peaks at smaller scales (larger multipoles) than the neutrino density contrast power spectrum because of the higher clustering of CDM. We have obtained very low auto- and cross-correlation spectra for the neutrino velocity modulus, as was also expected, because of the large thermal velocity of cosmic neutrinos with such a low mass. Regarding the deflection angles, we have obtained a larger signal in the cross-correlation of this variable with CDM density than in the cross-correlation between neutrino velocity and CDM density. 

The cross-correlations of neutrinos directly detected on Earth with CDM take into account the space-time delay between the arrival of neutrinos and photons. To enable meaningful cross-correlations, we considered photons carrying information from more distant structures, paired with neutrinos that were deflected by those same structures earlier in their paths, such that both arrive at Earth at the same time. Against naive expectations, these cross-correlations are nonzero, particularly at the largest scales for the neutrino overdensities and deflection angles. In addition, we have also explored an alternative indirect detection scenario, in which the neutrinos induce the emission of photons resulting in photons being detected on Earth simultaneously with those from large-scale structure, allowing for co-located cross-correlations. In this scenario, neutrino-induced photons loss the original neutrino directionality. However, despite the loss of angular information, the clustering scales of CDM and neutrinos remain consistent across both scenarios, validating the robustness of the signal on the largest scales.

The most directly related previous work to ours is that of Ref.~\cite{Elbers}. Note that while they focus on a constrained realization of the local Earth's neighborhood, we produce a statistical realization as we are interested on average properties as our aim was to study also the co-located signal, which is new. We also were interested in angle changes of the neutrino trajectory by large scale structure and the cross-correlation with the weak lensing signal. Because of this, we produced significantly larger ($2048^3$) simulations to study much lower resolution effects.

For the first time, cross-correlations between neutrinos and CDM maps have been shown in this paper, revealing the cosmological scales where any cross-correlation between them may occur. By demonstrating the angular dependence of these signals and identifying their maxima, this approach provides valuable insights into the nature of neutrinos through observations of the sky. Once the technology to measure these cosmological probes becomes available, examining these maps will enable us to confirm the detection of the cosmic neutrino background and facilitate its measurements as happens e.g. also for the integrated Sachs-Wolfe effect of CMB when exploiting its correlation with galaxies.

\acknowledgments 
The DEMNUni simulations were carried out in the framework of ``The Dark Energy and Massive Neutrino Universe" project, using the Tier-0 IBM BG/Q Fermi machine, the Tier-0 Intel OmniPath Cluster Marconi-A1 and Marconi-100 of the Centro Interuniversitario del Nord-Est per il Calcolo Elettronico (CINECA). CC acknowledges a generous CPU and storage allocation by the Italian Super-Computing Resource Allocation (ISCRA) as well as from the coordination of the ``Accordo Quadro MoU per lo svolgimento di attività congiunta di ricerca Nuove frontiere in Astrofisica: HPC e Data Exploration di nuova generazione'', together with storage from INFN-CNAF and INAF-IA2. This work was supported by the 'Center of Excellence Maria de Maeztu 2020-2023' award to ICCUB (CEX2019-000918-M) funded by MCIN/AEI/10.13039/501100011033. Funding for the work of RJ was partially provided by project PID2022-141125NB-I00. AG also thanks Martin Reinecke for many useful discussions about some technical aspects during the early stages of this work. MC is partially supported by the 2023/24 ``Research and Education'' grant from Fondazione CRT. The OAVdA is managed by the Fondazione Cl\'ement Fillietroz-ONLUS, which is supported by the Regional Government of the Aosta Valley, the Town Municipality of Nus and the ``Unit\'e des Communes vald\^otaines Mont-\'Emilius''.


\begin{thebibliography}{10}

\bibitem{Dolgov:2002wy}
A.~D. Dolgov, \emph{{Neutrinos in cosmology}}, \href{https://doi.org/10.1016/S0370-1573(02)00139-4}{\emph{Phys. Rept.} {\bfseries 370} (2002) 333} [\href{https://arxiv.org/abs/hep-ph/0202122}{{\ttfamily hep-ph/0202122}}].

\bibitem{Dolgov_2008}
A.~D. Dolgov, \emph{Cosmology and neutrino properties}, \href{https://doi.org/10.1134/s1063778808120181}{\emph{Physics of Atomic Nuclei} {\bfseries 71} (2008) 2152?2164} [\href{https://arxiv.org/abs/0803.3887}{{\ttfamily 0803.3887}}].

\bibitem{Long_2014}
A.~J. Long, C.~Lunardini and E.~Sabancilar, \emph{Detecting non-relativistic cosmic neutrinos by capture on tritium: phenomenology and physics potential}, \href{https://doi.org/10.1088/1475-7516/2014/08/038}{\emph{JCAP} {\bfseries 08} (2014) 038} [\href{https://arxiv.org/abs/1405.7654}{{\ttfamily 1405.7654}}].

\bibitem{PTolemy}
{\scshape PTOLEMY} collaboration, \emph{{Neutrino physics with the PTOLEMY project: active neutrino properties and the light sterile case}}, \href{https://doi.org/10.1088/1475-7516/2019/07/047}{\emph{JCAP} {\bfseries 07} (2019) 047} [\href{https://arxiv.org/abs/1902.05508}{{\ttfamily 1902.05508}}].

\bibitem{Pacos}
F.~Villaescusa-Navarro, S.~Bird, C.~Pena-Garay and M.~Viel, \emph{{Non-linear evolution of the cosmic neutrino background}}, \href{https://doi.org/10.1088/1475-7516/2013/03/019}{\emph{JCAP} {\bfseries 03} (2013) 019} [\href{https://arxiv.org/abs/1212.4855}{{\ttfamily 1212.4855}}].

\bibitem{Hernandez-Molinero_2024}
B.~Hernandez-Molinero, C.~Carbone, R.~Jimenez and C.~P. Garay, \emph{Cosmic background neutrinos deflected by gravity: Demnuni simulation analysis}, \href{https://doi.org/10.1088/1475-7516/2024/01/006}{\emph{Journal of Cosmology and Astroparticle Physics} {\bfseries 2024} (2024) 006}.

\bibitem{NS}
B.~Hernandez-Molinero, R.~Jimenez and C.~Pe\~na Garay, \emph{{Effects of the Cosmic Neutrino Background Capture on Astrophysical Objects}},  \href{https://arxiv.org/abs/2503.07788}{{\ttfamily 2503.07788}}.

\bibitem{DESI}
{DESI Collaboration}, A.~G. {Adame}, J.~{Aguilar}, S.~{Ahlen}, S.~{Alam}, D.~M. {Alexander} et~al., \emph{{DESI 2024 VI: Cosmological Constraints from the Measurements of Baryon Acoustic Oscillations}}, \href{https://doi.org/10.48550/arXiv.2404.03002}{\emph{arXiv e-prints} (2024) arXiv:2404.03002} [\href{https://arxiv.org/abs/2404.03002}{{\ttfamily 2404.03002}}].

\bibitem{Fer1}
F.~{Simpson}, R.~{Jimenez}, C.~{Pena-Garay} and L.~{Verde}, \emph{{Strong Bayesian evidence for the normal neutrino hierarchy}}, \href{https://doi.org/10.1088/1475-7516/2017/06/029}{\emph{\jcap} {\bfseries 2017} (2017) 029} [\href{https://arxiv.org/abs/1703.03425}{{\ttfamily 1703.03425}}].

\bibitem{Fer2}
R.~{Jimenez}, C.~{Pena-Garay}, K.~{Short}, F.~{Simpson} and L.~{Verde}, \emph{{Neutrino masses and mass hierarchy: evidence for the normal hierarchy}}, \href{https://doi.org/10.1088/1475-7516/2022/09/006}{\emph{\jcap} {\bfseries 2022} (2022) 006} [\href{https://arxiv.org/abs/2203.14247}{{\ttfamily 2203.14247}}].

\bibitem{Goeppert-Mayer}
M.~Goeppert-Mayer, \emph{Double beta-disintegration}, \href{https://doi.org/10.1103/PhysRev.48.512}{\emph{Phys. Rev.} {\bfseries 48} (1935) 512}.

\bibitem{furry}
W.~H. Furry, \emph{On transition probabilities in double beta-disintegration}, \href{https://doi.org/10.1103/PhysRev.56.1184}{\emph{Phys. Rev.} {\bfseries 56} (1939) 1184}.

\bibitem{Hernandez-Molinero:2022zoo}
B.~Hernandez-Molinero, R.~Jimenez and C.~Pena-Garay, \emph{{Distinguishing Dirac vs. Majorana neutrinos: a cosmological probe}}, \href{https://doi.org/10.1088/1475-7516/2022/08/038}{\emph{JCAP} {\bfseries 08} (2022) 038} [\href{https://arxiv.org/abs/2205.00808}{{\ttfamily 2205.00808}}].

\bibitem{Roulet:2018fyh}
E.~Roulet and F.~Vissani, \emph{{On the capture rates of big bang neutrinos by nuclei within the Dirac and Majorana hypotheses}}, \href{https://doi.org/10.1088/1475-7516/2018/10/049}{\emph{JCAP} {\bfseries 10} (2018) 049} [\href{https://arxiv.org/abs/1810.00505}{{\ttfamily 1810.00505}}].

\bibitem{Michney}
R.~J. Michney and R.~R. Caldwell, \emph{{Anisotropy of the Cosmic Neutrino Background}}, \href{https://doi.org/10.1088/1475-7516/2007/01/014}{\emph{JCAP} {\bfseries 01} (2007) 014} [\href{https://arxiv.org/abs/astro-ph/0608303}{{\ttfamily astro-ph/0608303}}].

\bibitem{Hannestad}
S.~{Hannestad} and J.~{Brandbyge}, \emph{{The Cosmic Neutrino Background anisotropy {\textemdash} linear theory}}, \href{https://doi.org/10.1088/1475-7516/2010/03/020}{\emph{\jcap} {\bfseries 2010} (2010) 020} [\href{https://arxiv.org/abs/0910.4578}{{\ttfamily 0910.4578}}].

\bibitem{Tully}
C.~G. {Tully} and G.~{Zhang}, \emph{{Multi-messenger astrophysics with the cosmic neutrino background}}, \href{https://doi.org/10.1088/1475-7516/2021/06/053}{\emph{\jcap} {\bfseries 2021} (2021) 053} [\href{https://arxiv.org/abs/2103.01274}{{\ttfamily 2103.01274}}].

\bibitem{Elbers}
W.~{Elbers}, C.~S. {Frenk}, A.~{Jenkins}, B.~{Li}, S.~{Pascoli}, J.~{Jasche} et~al., \emph{{Where shadows lie: reconstruction of anisotropies in the neutrino sky}}, \href{https://doi.org/10.1088/1475-7516/2023/10/010}{\emph{\jcap} {\bfseries 2023} (2023) 010} [\href{https://arxiv.org/abs/2307.03191}{{\ttfamily 2307.03191}}].

\bibitem{carbone_2016}
C.~{Carbone}, M.~{Petkova} and K.~{Dolag}, \emph{{DEMNUni: ISW, Rees-Sciama, and weak-lensing in the presence of massive neutrinos}}, \href{https://doi.org/10.1088/1475-7516/2016/07/034}{\emph{\jcap} {\bfseries 2016} (2016) 034} [\href{https://arxiv.org/abs/1605.02024}{{\ttfamily 1605.02024}}].

\bibitem{castorina_2015}
E.~{Castorina}, C.~{Carbone}, J.~{Bel}, E.~{Sefusatti} and K.~{Dolag}, \emph{{DEMNUni: the clustering of large-scale structures in the presence of massive neutrinos}}, \href{https://doi.org/10.1088/1475-7516/2015/07/043}{\emph{\jcap} {\bfseries 7} (2015) 043} [\href{https://arxiv.org/abs/1505.07148}{{\ttfamily 1505.07148}}].

\bibitem{moresco_2016}
M.~{Moresco}, F.~{Marulli}, L.~{Moscardini}, E.~{Branchini}, A.~{Cappi}, I.~{Davidzon} et~al., \emph{{The VIMOS Public Extragalactic Redshift Survey (VIPERS) . Exploring the dependence of the three-point correlation function on stellar mass and luminosity at 0.5 <z < 1.1}}, \href{https://doi.org/10.1051/0004-6361/201628589}{\emph{\aap} {\bfseries 604} (2017) A133} [\href{https://arxiv.org/abs/1603.08924}{{\ttfamily 1603.08924}}].

\bibitem{zennaro_2018}
M.~{Zennaro}, J.~{Bel}, J.~{Dossett}, C.~{Carbone} and L.~{Guzzo}, \emph{{Cosmological constraints from galaxy clustering in the presence of massive neutrinos}}, \href{https://doi.org/10.1093/mnras/sty670}{\emph{\mnras} {\bfseries 477} (2018) 491} [\href{https://arxiv.org/abs/1712.02886}{{\ttfamily 1712.02886}}].

\bibitem{ruggeri_2018}
R.~{Ruggeri}, E.~{Castorina}, C.~{Carbone} and E.~{Sefusatti}, \emph{{DEMNUni: massive neutrinos and the bispectrum of large scale structures}}, \href{https://doi.org/10.1088/1475-7516/2018/03/003}{\emph{\jcap} {\bfseries 2018} (2018) 003} [\href{https://arxiv.org/abs/1712.02334}{{\ttfamily 1712.02334}}].

\bibitem{bel_2019}
J.~{Bel}, A.~{Pezzotta}, C.~{Carbone}, E.~{Sefusatti} and L.~{Guzzo}, \emph{{Accurate fitting functions for peculiar velocity spectra in standard and massive-neutrino cosmologies}}, \href{https://doi.org/10.1051/0004-6361/201834513}{\emph{\aap} {\bfseries 622} (2019) A109} [\href{https://arxiv.org/abs/1809.09338}{{\ttfamily 1809.09338}}].

\bibitem{parimbelli_2021}
G.~{Parimbelli}, S.~{Anselmi}, M.~{Viel}, C.~{Carbone}, F.~{Villaescusa-Navarro}, P.~S. {Corasaniti} et~al., \emph{{The effects of massive neutrinos on the linear point of the correlation function}}, \href{https://doi.org/10.1088/1475-7516/2021/01/009}{\emph{\jcap} {\bfseries 2021} (2021) 009} [\href{https://arxiv.org/abs/2007.10345}{{\ttfamily 2007.10345}}].

\bibitem{parimbelli_2022}
G.~{Parimbelli}, C.~{Carbone}, J.~{Bel}, B.~{Bose}, M.~{Calabrese}, E.~{Carella} et~al., \emph{{DEMNUni: comparing nonlinear power spectra prescriptions in the presence of massive neutrinos and dynamical dark energy}}, \href{https://doi.org/10.1088/1475-7516/2022/11/041}{\emph{\jcap} {\bfseries 2022} (2022) 041} [\href{https://arxiv.org/abs/2207.13677}{{\ttfamily 2207.13677}}].

\bibitem{Guidi_2022}
M.~{Guidi}, A.~{Veropalumbo}, E.~{Branchini}, A.~{Eggemeier} and C.~{Carbone}, \emph{{Modelling the next-to-leading order matter three-point correlation function using FFTLog}}, \href{https://doi.org/10.1088/1475-7516/2023/08/066}{\emph{\jcap} {\bfseries 2023} (2023) 066} [\href{https://arxiv.org/abs/2212.07382}{{\ttfamily 2212.07382}}].

\bibitem{Baratta_2022}
P.~{Baratta}, J.~{Bel}, S.~{Gouyou Beauchamps} and C.~{Carbone}, \emph{{COVMOS: a new Monte Carlo approach for galaxy clustering analysis}}, \href{https://doi.org/10.48550/arXiv.2211.13590}{\emph{arXiv e-prints} (2022) arXiv:2211.13590} [\href{https://arxiv.org/abs/2211.13590}{{\ttfamily 2211.13590}}].

\bibitem{Gouyou_Beauchamps_2023}
S.~{Gouyou Beauchamps}, P.~{Baratta}, S.~{Escoffier}, W.~{Gillard}, J.~{Bel}, J.~{Bautista} et~al., \emph{{Cosmological inference including massive neutrinos from the matter power spectrum: biases induced by uncertainties in the covariance matrix}}, \href{https://doi.org/10.48550/arXiv.2306.05988}{\emph{arXiv e-prints} (2023) arXiv:2306.05988} [\href{https://arxiv.org/abs/2306.05988}{{\ttfamily 2306.05988}}].

\bibitem{Hernandez_2024b}
B.~{Hern{\'a}ndez-Molinero}, C.~{Carbone}, R.~{Jimenez} and C.~P. {Garay}, \emph{{Neutrino halo profiles: HR-DEMNUni simulation analysis}}, \href{https://doi.org/10.1088/1475-7516/2024/09/033}{\emph{\jcap} {\bfseries 2024} (2024) 033} [\href{https://arxiv.org/abs/2407.12694}{{\ttfamily 2407.12694}}].

\bibitem{Verdiani_2025}
F.~{Verdiani}, E.~{Bellini}, C.~{Moretti}, E.~{Sefusatti}, C.~{Carbone} and M.~{Viel}, \emph{{Redshift-Space Distortions in Massive Neutrinos Cosmologies}}, \href{https://doi.org/10.48550/arXiv.2503.06655}{\emph{arXiv e-prints} (2025) arXiv:2503.06655} [\href{https://arxiv.org/abs/2503.06655}{{\ttfamily 2503.06655}}].

\bibitem{SHAM-Carella_in_prep}
E.~{Carella}, C.~{Carbone}, M.~{Zennaro}, G.~{Girelli}, M.~{Bolzonella}, F.~{Marulli} et~al., ``Demnuni: The galaxy-halo connection in the presence of dynamical dark energy and massive neutrinos.''.

\bibitem{roncarelli_2015}
M.~{Roncarelli}, C.~{Carbone} and L.~{Moscardini}, \emph{{The effect of massive neutrinos on the Sunyaev-Zel'dovich and X-ray observables of galaxy clusters}}, \href{https://doi.org/10.1093/mnras/stu2546}{\emph{\mnras} {\bfseries 447} (2015) 1761} [\href{https://arxiv.org/abs/1409.4285}{{\ttfamily 1409.4285}}].

\bibitem{fabbian_2018}
G.~{Fabbian}, M.~{Calabrese} and C.~{Carbone}, \emph{{CMB weak-lensing beyond the Born approximation: a numerical approach}}, \href{https://doi.org/10.1088/1475-7516/2018/02/050}{\emph{\jcap} {\bfseries 2018} (2018) 050} [\href{https://arxiv.org/abs/1702.03317}{{\ttfamily 1702.03317}}].

\bibitem{Ingoglia_2024}
{Euclid Collaboration}, L.~{Ingoglia}, M.~{Sereno}, S.~{Farrens}, C.~{Giocoli}, L.~{Baumont} et~al., \emph{{Euclid preparation: Determining the weak lensing mass accuracy and precision for galaxy clusters}}, \href{https://doi.org/10.48550/arXiv.2409.02783}{\emph{arXiv e-prints} (2024) arXiv:2409.02783} [\href{https://arxiv.org/abs/2409.02783}{{\ttfamily 2409.02783}}].

\bibitem{Luchina_etal_2025}
D.~{Luchina}, M.~{Roncarelli}, M.~{Calabrese}, G.~{Fabbian} and C.~{Carbone}, \emph{{DEMNUni: the Sunyaev-Zel'dovich effect in the presence of massive neutrinos and dynamical dark energy}}, \href{https://doi.org/10.48550/arXiv.2503.16355}{\emph{arXiv e-prints} (2025) arXiv:2503.16355} [\href{https://arxiv.org/abs/2503.16355}{{\ttfamily 2503.16355}}].

\bibitem{kreisch_2019}
C.~D. {Kreisch}, A.~{Pisani}, C.~{Carbone}, J.~{Liu}, A.~J. {Hawken}, E.~{Massara} et~al., \emph{{Massive neutrinos leave fingerprints on cosmic voids}}, \href{https://doi.org/10.1093/mnras/stz1944}{\emph{\mnras} {\bfseries 488} (2019) 4413} [\href{https://arxiv.org/abs/1808.07464}{{\ttfamily 1808.07464}}].

\bibitem{schuster_2019}
N.~{Schuster}, N.~{Hamaus}, A.~{Pisani}, C.~{Carbone}, C.~D. {Kreisch}, G.~{Pollina} et~al., \emph{{The bias of cosmic voids in the presence of massive neutrinos}}, \href{https://doi.org/10.1088/1475-7516/2019/12/055}{\emph{\jcap} {\bfseries 2019} (2019) 055} [\href{https://arxiv.org/abs/1905.00436}{{\ttfamily 1905.00436}}].

\bibitem{Verza_2019}
G.~{Verza}, A.~{Pisani}, C.~{Carbone}, N.~{Hamaus} and L.~{Guzzo}, \emph{{The void size function in dynamical dark energy cosmologies}}, \href{https://doi.org/10.1088/1475-7516/2019/12/040}{\emph{\jcap} {\bfseries 2019} (2019) 040} [\href{https://arxiv.org/abs/1906.00409}{{\ttfamily 1906.00409}}].

\bibitem{Verza_2022a}
G.~{Verza}, C.~{Carbone} and A.~{Renzi}, \emph{{The Halo Bias inside Cosmic Voids}}, \href{https://doi.org/10.3847/2041-8213/ac9d98}{\emph{\apjl} {\bfseries 940} (2022) L16} [\href{https://arxiv.org/abs/2207.04039}{{\ttfamily 2207.04039}}].

\bibitem{Verza_2022b}
G.~{Verza}, C.~{Carbone}, A.~{Pisani} and A.~{Renzi}, \emph{{DEMNUni: disentangling dark energy from massive neutrinos with the void size function}}, \href{https://doi.org/10.48550/arXiv.2212.09740}{\emph{arXiv e-prints} (2022) arXiv:2212.09740} [\href{https://arxiv.org/abs/2212.09740}{{\ttfamily 2212.09740}}].

\bibitem{Verza_2024}
G.~{Verza}, C.~{Carbone}, A.~{Pisani}, C.~{Porciani} and S.~{Matarrese}, \emph{{The universal multiplicity function: counting halos and voids}}, \href{https://doi.org/10.48550/arXiv.2401.14451}{\emph{arXiv e-prints} (2024) arXiv:2401.14451} [\href{https://arxiv.org/abs/2401.14451}{{\ttfamily 2401.14451}}].

\bibitem{Vielzeuf_2022}
P.~{Vielzeuf}, M.~{Calabrese}, C.~{Carbone}, G.~{Fabbian} and C.~{Baccigalupi}, \emph{{DEMNUni: the imprint of massive neutrinos on the cross-correlation between cosmic voids and CMB lensing}}, \href{https://doi.org/10.1088/1475-7516/2023/08/010}{\emph{\jcap} {\bfseries 2023} (2023) 010} [\href{https://arxiv.org/abs/2303.10048}{{\ttfamily 2303.10048}}].

\bibitem{Cuozzo2022}
V.~{Cuozzo}, C.~{Carbone}, M.~{Calabrese}, E.~{Carella} and M.~{Migliaccio}, \emph{{DEMNUni: cross-correlating the nonlinear ISWRS effect with CMB-lensing and galaxies in the presence of massive neutrinos}}, \href{https://doi.org/10.48550/arXiv.2307.15711}{\emph{arXiv e-prints} (2023) arXiv:2307.15711} [\href{https://arxiv.org/abs/2307.15711}{{\ttfamily 2307.15711}}].

\bibitem{zennaro_2017}
M.~{Zennaro}, J.~{Bel}, F.~{Villaescusa-Navarro}, C.~{Carbone}, E.~{Sefusatti} and L.~{Guzzo}, \emph{{Initial conditions for accurate N-body simulations of massive neutrino cosmologies}}, \href{https://doi.org/10.1093/mnras/stw3340}{\emph{\mnras} {\bfseries 466} (2017) 3244} [\href{https://arxiv.org/abs/1605.05283}{{\ttfamily 1605.05283}}].

\bibitem{planck2013}
{Planck Collaboration}, P.~A.~R. {Ade}, N.~{Aghanim}, C.~{Armitage-Caplan}, M.~{Arnaud}, M.~{Ashdown} et~al., \emph{{Planck 2013 results. XVI. Cosmological parameters}}, \href{https://doi.org/10.1051/0004-6361/201321591}{\emph{\aap} {\bfseries 571} (2014) A16} [\href{https://arxiv.org/abs/1303.5076}{{\ttfamily 1303.5076}}].

\bibitem{Gorski}
K.~M. {G{\'o}rski}, E.~{Hivon}, A.~J. {Banday}, B.~D. {Wandelt}, F.~K. {Hansen}, M.~{Reinecke} et~al., \emph{{HEALPix: A Framework for High-Resolution Discretization and Fast Analysis of Data Distributed on the Sphere}}, \href{https://doi.org/10.1086/427976}{\emph{\apj} {\bfseries 622} (2005) 759} [\href{https://arxiv.org/abs/astro-ph/0409513}{{\ttfamily astro-ph/0409513}}].

\bibitem{carbone2009}
C.~{Carbone}, C.~{Baccigalupi}, M.~{Bartelmann}, S.~{Matarrese} and V.~{Springel}, \emph{{Lensed CMB temperature and polarization maps from the Millennium Simulation}}, \href{https://doi.org/10.1111/j.1365-2966.2009.14746.x}{\emph{\mnras} {\bfseries 396} (2009) 668} [\href{https://arxiv.org/abs/0810.4145}{{\ttfamily 0810.4145}}].

\bibitem{calabrese2015}
M.~{Calabrese}, C.~{Carbone}, G.~{Fabbian}, M.~{Baldi} and C.~{Baccigalupi}, \emph{{Multiple lensing of the cosmic microwave background anisotropies}}, \href{https://doi.org/10.1088/1475-7516/2015/03/049}{\emph{\jcap} {\bfseries 3} (2015) 049} [\href{https://arxiv.org/abs/1409.7680}{{\ttfamily 1409.7680}}].

\bibitem{fabbian2018}
G.~{Fabbian}, M.~{Calabrese} and C.~{Carbone}, \emph{{CMB weak-lensing beyond the Born approximation: a numerical approach}}, \href{https://doi.org/10.1088/1475-7516/2018/02/050}{\emph{\jcap} {\bfseries 2018} (2018) 050} [\href{https://arxiv.org/abs/1702.03317}{{\ttfamily 1702.03317}}].

\bibitem{Accuracy_weak_lensing_simulations-Hilbert+20}
S.~{Hilbert}, A.~{Barreira}, G.~{Fabbian}, P.~{Fosalba}, C.~{Giocoli}, S.~{Bose} et~al., \emph{{The accuracy of weak lensing simulations}}, \href{https://doi.org/10.1093/mnras/staa281}{\emph{\mnras} {\bfseries 493} (2020) 305} [\href{https://arxiv.org/abs/1910.10625}{{\ttfamily 1910.10625}}].

\bibitem{Laureijs_2011}
R.~{Laureijs}, J.~{Amiaux}, S.~{Arduini}, J.~L. {Augu{\`e}res}, J.~{Brinchmann}, R.~{Cole} et~al., \emph{{Euclid Definition Study Report}}, \href{https://doi.org/10.48550/arXiv.1110.3193}{\emph{arXiv e-prints} (2011) arXiv:1110.3193} [\href{https://arxiv.org/abs/1110.3193}{{\ttfamily 1110.3193}}].

\bibitem{EuclidVII_2020}
{Euclid Collaboration}, A.~{Blanchard}, S.~{Camera}, C.~{Carbone}, V.~F. {Cardone}, S.~{Casas} et~al., \emph{{Euclid preparation. VII. Forecast validation for Euclid cosmological probes}}, \href{https://doi.org/10.1051/0004-6361/202038071}{\emph{\aap} {\bfseries 642} (2020) A191} [\href{https://arxiv.org/abs/1910.09273}{{\ttfamily 1910.09273}}].

\bibitem{2001PhR...340..291B}
M.~{Bartelmann} and P.~{Schneider}, \emph{{Weak gravitational lensing}}, \href{https://doi.org/10.1016/S0370-1573(00)00082-X}{\emph{\physrep} {\bfseries 340} (2001) 291} [\href{https://arxiv.org/abs/astro-ph/9912508}{{\ttfamily astro-ph/9912508}}].

\end{thebibliography}
\bibliographystyle{JHEP}

\providecommand{\href}[2]{#2}\begingroup\raggedright\endgroup

\appendix
\section{Methodology to estimate the past neutrino cone at low redshift.}
\label{sec:past_neutrino_methods}

To further clarify our method regarding N-body simulations and the construction of a pastcone for neutrinos, and to show that for the last snapshot, integrating over the box or doing shells is equivalent, we conducted the following test and comparison. We constructed a full past cone by placing an observer at the center of the universe, at redshift $z=0$, and generating a full sky cone extending backward in time. This was achieved using the appropriate snapshots from the simulation of the N body at $z=0,0.0135829,0.0319811,0.0602070$ (z = 0.06 corresponds to a comoving distance of 250 Mpc h$^{-1}$), as the redshift changes while tracking in time. For this purpose, we used the above N-body snapshots, which accurately track the evolution of neutrinos and dark matter from $z=0$ to the redshift corresponding to a distance of 250 Mpc$h^{-1}$.

For this light-cone, we created a full-sky map of both dark-matter and neutrino surface mass density. The particles in the simulation from each shell of the lightcone were projected onto a two-dimensional HEALPix map. This procedure is widely used in the literature to construct weak lensing observables.

However, this construction of past light cones was not used in the manuscript. As mentioned above, for this work, we considered a single snapshot at $z=0$, taking into account all dark matter and neutrino particles within a radius of 250 Mpc$h^{-1}$ around the observer, and projecting all matter onto a single surface mass density map. It is important to note that this is indeed an approximation, which we employed to cross-correlate the mass densities of dark matter and neutrinos with neutrino features such as their deflection angles and velocities.

\begin{figure}
    \centering
    \begin{subfigure}[b]{0.48\linewidth}
        \centering
    \includegraphics[width=\linewidth]{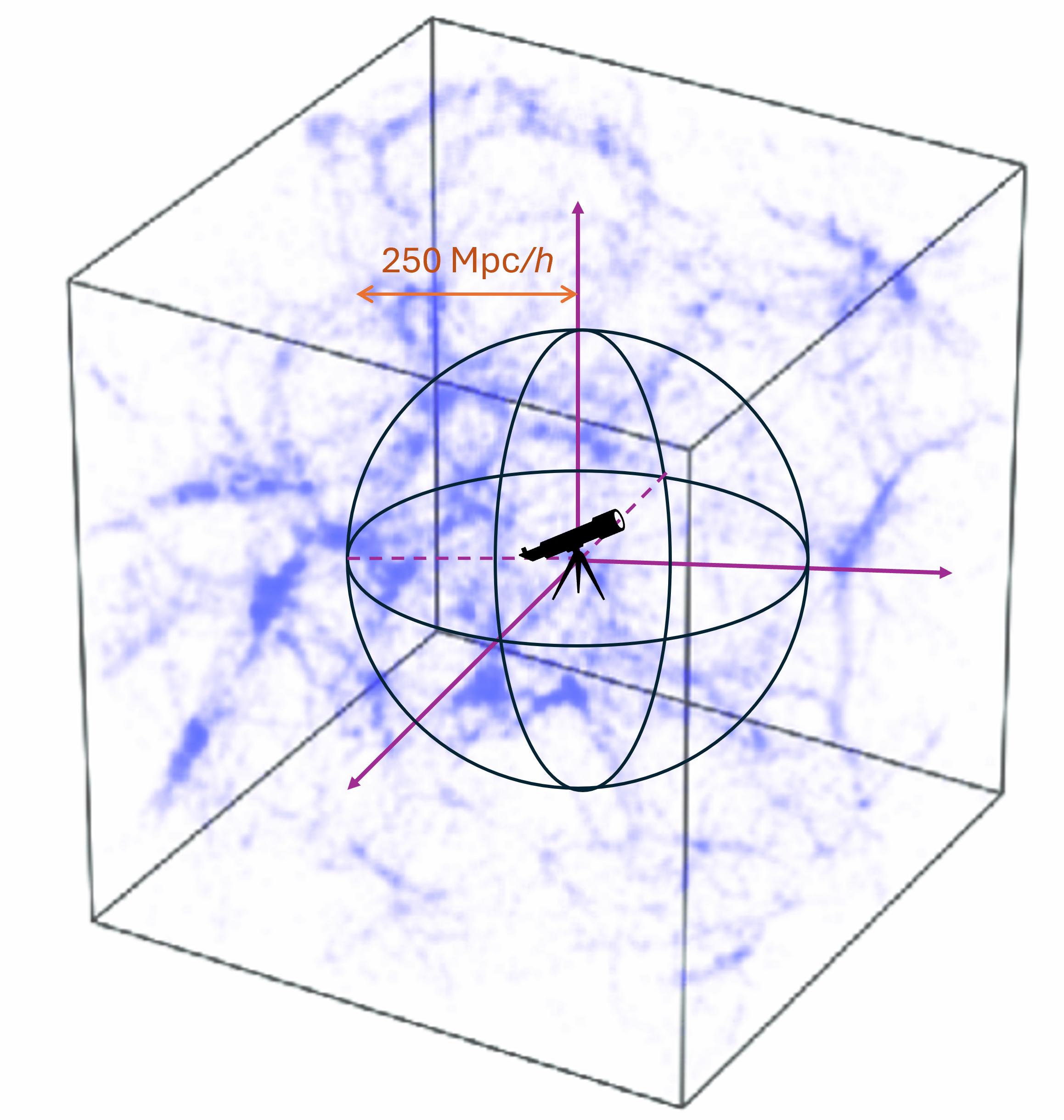}
        \caption{Single snapshot, $z=0$}
        \label{fig:map_proc_250Mpc}
    \end{subfigure}
    \hfill
    \begin{subfigure}[b]{0.48\linewidth}
        \centering
    \includegraphics[width=\linewidth]{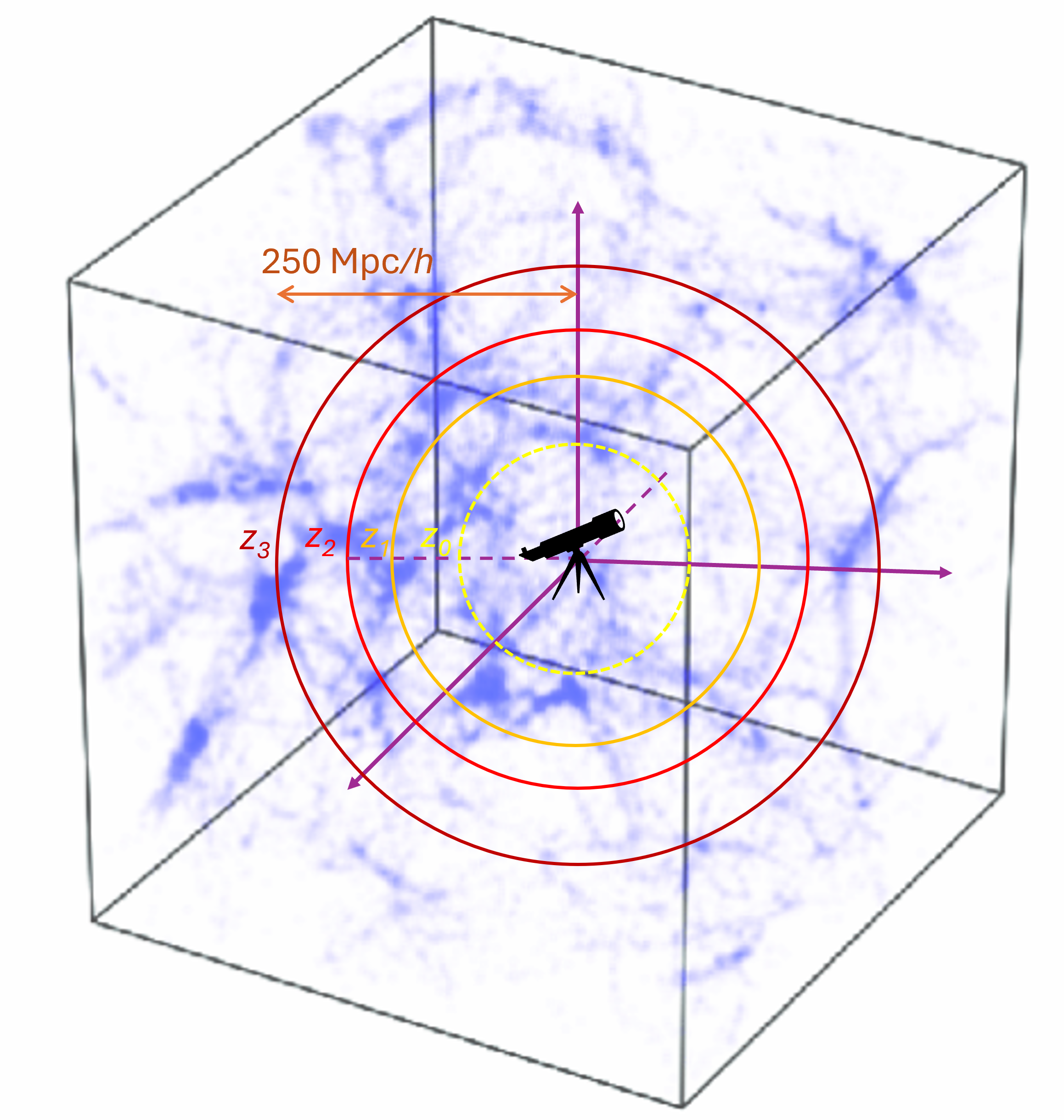}
        \caption{Full past-lightcone}
        \label{fig:map_proc_lightcone}
    \end{subfigure}
    \caption{Comparison between the two different map procedures, as described in the text. In this work we have used the lighcone procedure drawn in \autoref{fig:map_proc_250Mpc}.}
    \label{fig:map_procedure}
\end{figure}

\begin{figure}
\centering
\begin{tabular}{cc}
\includegraphics[width=0.49\textwidth]{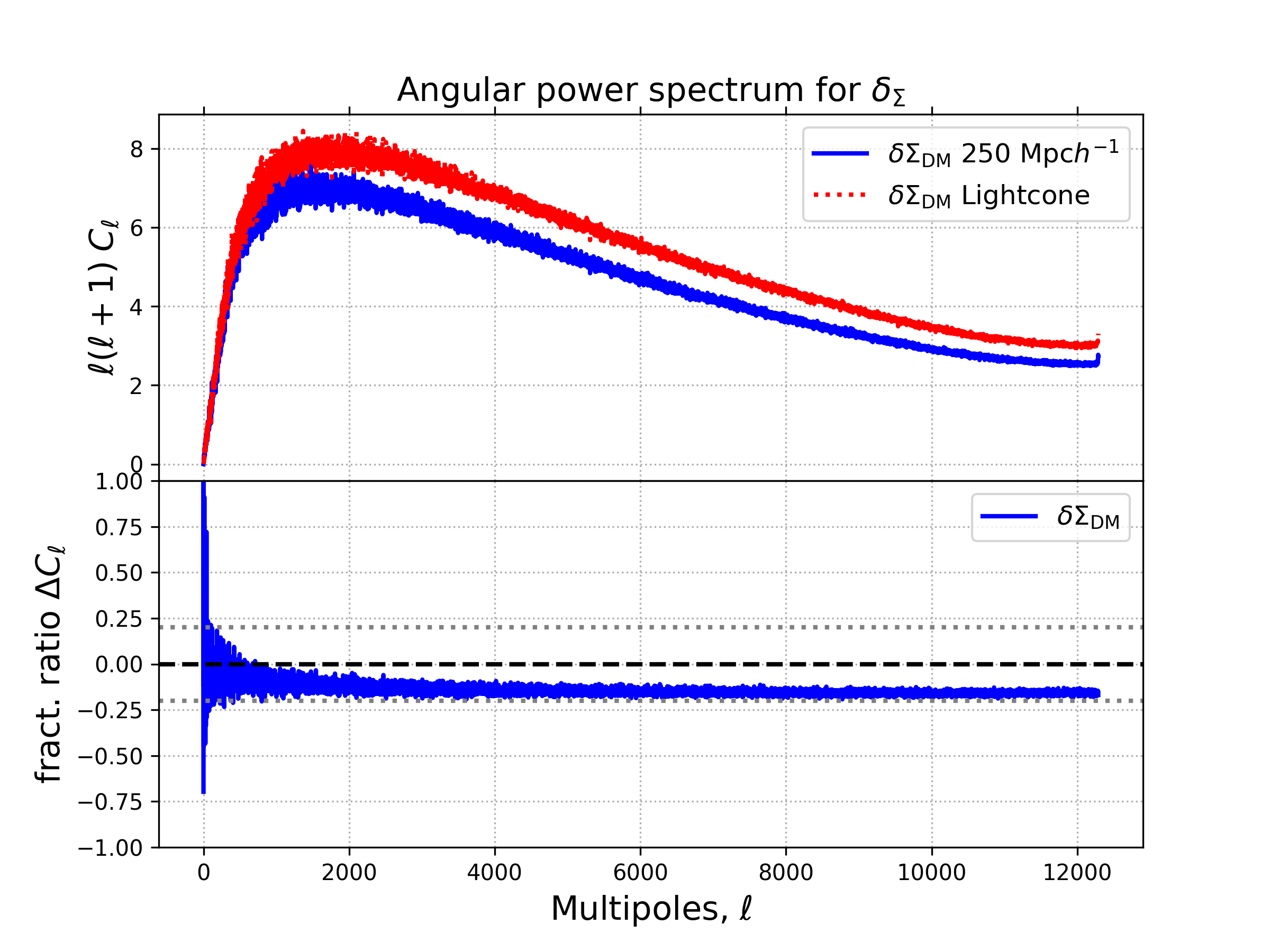}
\includegraphics[width=0.49\textwidth]{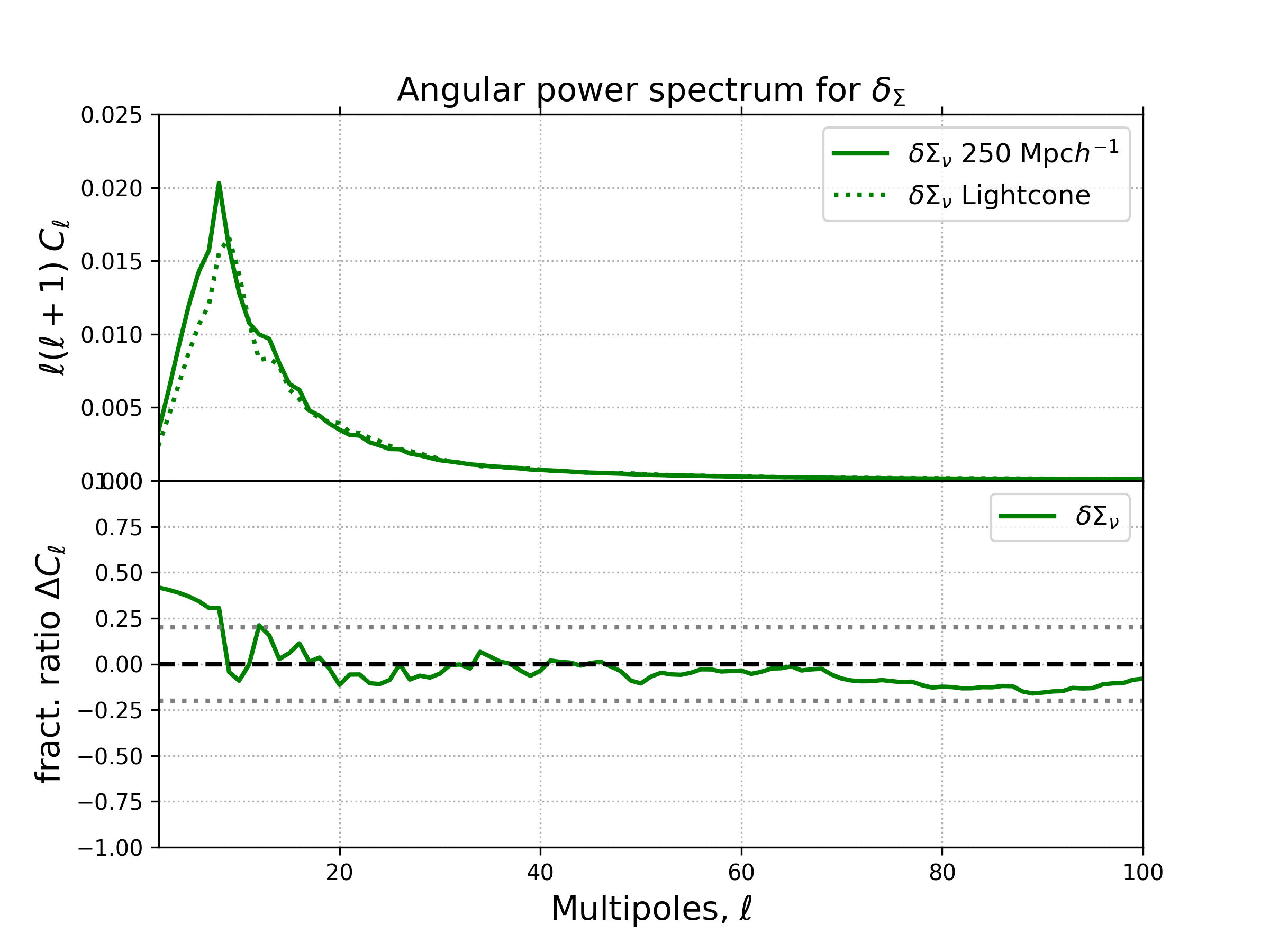}
\end{tabular}
\caption{Angular power spectrum comparison between the two different map-making procedures described in the text. The sub-plots of the left and right panels show the fractional ratio, $\Delta C_\ell$, between the angular power spectra extracted from the one-snapshot map and the full past-cone.}
\label{fig:comparison}
\end{figure}

In \autoref{fig:map_procedure}, we provide a sketch of these two different procedures for clarity. Furthermore, in \autoref{fig:comparison}, we present a comparison of the angular power spectra derived from the surface mass density contrast $\delta_{\Sigma}$ for the two lightcone constructions: the complete past-lightcone reconstruction and the single snapshot at $z=0$. The two panels show the comparison for the dark matter and neutrino components, respectively.

As expected, the two procedures yield similar results; differences - particularly for neutrinos - are minor and at the percentage level. This demonstrates that, although our approach is an approximation, as stated in the paper, it is a reasonable one, capable of capturing the matter distribution and adequately describing the neutrino cross-correlation.

Our presented technique can be useful for computing neutrino properties with mass $< 0.1$ eV.

\end{document}